\documentclass[11pt]{article}
\pdfoutput=1
\usepackage[margin=1in]{geometry}
\usepackage{amsmath, amssymb, braket, dsfont,authblk}
\usepackage{graphicx}
\usepackage{rotating, mathtools}
\usepackage[matrix,arrow]{xy}
\usepackage[numbers]{natbib}
\usepackage[colorlinks]{hyperref}
\usepackage{arydshln}
\usepackage{caption}

\numberwithin{equation}{section}

\hypersetup{linkcolor={blue},citecolor={purple}}
\usepackage{tikz}
\usepackage{listings}
\usetikzlibrary{calc}
 \usetikzlibrary{decorations}
\usetikzlibrary{plotmarks}

\newcommand{\comment}[1]{}

\newcommand{\zn}{{\mathbb{Z}_n}}
\newcommand{\be}{\begin{equation}}
\newcommand{\ee}{\end{equation}}
\newcommand{\bea}{\begin{eqnarray}}
\newcommand{\eea}{\end{eqnarray}}

\newcommand{\Qh}{\widehat{Q}}

\newcommand{\Ss}{\mathbf{S}}

\title{Onsager symmetries in $U(1)$-invariant clock models}
\author[1]{Eric Vernier}
\author[1]{Edward O'Brien}
\author[1,2]{Paul Fendley}
\affil[1]{\small Rudolf Peierls Centre for Theoretical Physics, Parks Rd, Oxford OX1 3PU, United Kingdom}
\affil[2]{\small All Souls College, Oxford, OX1 4AL, United Kingdom}

\begin{document}
\maketitle

\begin{abstract}

We show how the Onsager algebra, used in the original solution of the two-dimensional Ising model, arises as an infinite-dimensional symmetry of certain self-dual models that also have a $U(1)$ symmetry. We 
describe in detail the example of nearest-neighbour $n$-state clock chains whose ${\mathbb Z}_n$ symmetry is enhanced to $U(1)$.  As a consequence of the Onsager-algebra symmetry, the spectrum of these models  possesses degeneracies with multiplicities $2^N$ for positive integer $N$. We construct the elements of the algebra explicitly from transfer matrices built from non-fundamental representations of the quantum-group algebra $U_q(sl_2)$. We analyse the spectra further by using both the coordinate Bethe ansatz and a functional approach, and show that the degeneracies result from special exact $n$-string solutions of the Bethe equations. We also find a family of commuting chiral Hamiltonians that break the degeneracies and allow an integrable interpolation between ferro- and antiferromagnets. 

\end{abstract}

\section{Introduction}

One of the great triumphs of theoretical physics is Onsager's solution of the two-dimensional Ising model of statistical mechanics \cite{Onsager44}. As his solution included the case of anisotropic couplings, it yields also the spectrum of the associated Ising quantum spin chain. It is therefore quite surprising that his method of solution is not widely known, and seems to have gained a reputation for incomprehensibility. This is rather a shame, since not only is the calculation quite clear and elegant,\footnote{The reputation likely arose because Onsager wrote out the details of the calculation, instead of following the currently fashionable practice of treating essential knowledge as supplemental information.} but he describes a very interesting approach to the problem.

What Onsager did was to show that the transfer matrix can be constructed in terms of operators that are elements of a very elegant and simple infinite-dimensional Lie algebra now bearing his name. Other elements of the Onsager algebra are not part of the transfer matrix, but still have nice commutation properties. These other operators thus can be used to construct raising and lowering operators that map between eigenstates of the transfer matrix or quantum Hamiltonian with different energies. The exact spectrum then can be computed by exploiting one further property: the particular representation of the algebra arising in the Ising model is finite-dimensional, with size depending only linearly on the number of sites. Namely, with periodic boundary conditions, the elements of this representation themselves obey a nice periodicity condition, allowing quasiparticle momenta to be defined and quantised.

Soon after Onsager's work, Kaufman realised \cite{Kaufman49a,Kaufman49b} that fermionic operators arise naturally in the Ising model, allowing a closely related but distinct approach to solving the model using a Jordan-Wigner transformation \cite{Schultz64}. This fermionic method is even easier, and so for the most part the Onsager algebra itself was no longer exploited. Moreover, the elements of the Onsager algebra in the Ising model are all free-fermion bilinears, and their commutation relations are nice because any commutator of such fermion bilinears yields a linear combination of bilinears. 

It thus seemed sensible to expect that the Onsager algebra is merely one of the many marvellous properties of free-fermionic systems, and so only occurs in such. This expectation, however, is simply wrong. Motivated by some curious observations by Howes, Kadanoff and den Nijs \cite{Howes83}, von Gehlen and Rittenberg made the remarkable observation that the Onsager algebra is obeyed by operators in an $n$-state clock model \cite{Gehlen84}. They then show that the algebra allows construction of an infinite series of commuting local conserved charges, strongly suggesting a certain chiral clock model commuting with them is integrable.\footnote{Ironically, the Onsager algebra here does not shed much light on the original curious observations of \cite{Howes83}, which instead are best understood by utilising parafermionic operators \cite{Fradkin80,Fendley12}.} This model is indeed integrable, and now bears the name of the superintegrable chiral Potts model. The integrability allows many of its properties to be computed \cite{Albertini89}, but the Onsager algebra is not heavily utilised in this analysis. 

The Onsager algebra cannot be used to solve chiral clock models directly because its elements here do not have the simple periodicity property that the Ising presentation has. Thus what is free-fermionic about Onsager's original solution is the periodicity property, not the algebra itself. However, while progress has been made in understanding how the Onsager algebra relates to more standard approaches to integrability (see \cite{Baseilhac18} and references therein), the question remains: what more can the Onsager algebra tell us about properties of clock models? 

The purpose of this paper is to define and analyse a series of clock models that have the Onsager algebra as a symmetry algebra: all its elements commute with the Hamiltonian and transfer matrix. We believe this is the simplest set of such models. We dub them the self-dual $U(1)$ clock models, as the ${\mathbb Z}_n$ symmetry is promoted to a full $U(1)$ here. Although these models turn out to be special cases of the integrable XXZ chains of higher spin, the Onsager-algebra symmetry results in a number of striking properties not well understood in the more general setting. In particular, the spectrum should contain degeneracies because this symmetry algebra is non-abelian. We show that for any $n$, these degeneracies do appear, organising the states into multiplets of size $2^N$ for integer $N$. More general degeneracies in the XXZ spectrum have been found by using a ``loop-group'' symmetry \cite{Korff01,FabriciusMcCoy,FabriciusMcCoy2,NishinoDeguchi}, but the approach here is much simpler. Indeed, our results are quite reminiscent of the appearance of Yangian symmetries in long-range quantum-spin chains \cite{Haldane94}. 

The basic idea behind our approach is to exploit the combination of self-duality with $U(1)$ symmetry. Kramers-Wannier duality originally arose in the Ising model, relating a partition function in the disordered phase to one in the ordered, with the phase transition occurring at the self-dual coupling \cite{Kramers41}. One of the main motivations for introducing clock models with ${\mathbb Z}_n$ symmetry and Potts models with $S_n$ symmetry was to give other models exhibiting the same type of duality \cite{Baxter82}. We consider nearest-neighbour self-dual clock models whose Hamiltonians and transfer matrices preserve a $U(1)$ symmetry.  The self-duality means that the models must exhibit {\em two} $U(1)$ symmetries, the original one generated by an operator $Q$, and another one generator by its dual $\Qh$. 

The key observation is that these two $U(1)$ symmetry operators do not commute, but in fact generate the Onsager algebra! This proves remarkably easy to see. Namely, in a nearest-neighbour Hamiltonian such as ours, acting with $\Qh$ can change the eigenvalues of $Q$ only by $0,\pm n$. We show this fact explicitly below in section \ref{sec:Onsager}. Thus the dual $U(1)$ operator can be decomposed into a sum of three terms as
\be 
\widehat{Q} = {Q}^0  + {Q}^+  + {Q}^- \,, 
\label{qhatdecomposition}
\ee 
where $Q^\pm$ change the charge by $\pm n$, so that
\begin{align}
\big[Q,\,{Q}^0\big]=0\ ,\qquad\quad \big[Q,\, {Q}^\pm] = \pm n\,{Q}^\pm \ .
\label{Qpmalg}
\end{align}
Because $\widehat{Q}$ can be decomposed in such a fashion, it follows immediately that 
\begin{align*}
\big[Q,\,\widehat{Q}\big] &= n\big(Q^+-Q^-\big)\ ,\cr
\big[Q,\big[Q,\,\widehat{Q}\big]\big] &= n^2\big(Q^+ +Q^-\big)\ ,\cr
\Big[Q,\big[Q,\big[Q,\,\widehat{Q}\big]\big]\Big] &= n^3\big(Q^+ -Q^-\big)\ .
\end{align*}
Therefore
\begin{align}
\Big[Q,\big[Q,\big[Q,\,\widehat{Q}\big]\big]\Big]= n^2 \big[Q,\,\widehat{Q}\big]\ ,
\label{Dolan1}
\end{align}
and then self-duality requires
\begin{align}
\Big[\Qh,\big[\Qh,\big[\Qh,\,Q\big]\big]\Big] =n^2 \big[\widehat{Q},\,Q\big]\ .
\label{Dolan2}
\end{align}

The relations \eqref{Dolan1} and \eqref{Dolan2} are known as the {\em Dolan-Grady relations} \cite{Dolan81}. Repeatedly commuting with $Q$ and $\Qh$ subject to these constraints generates the Onsager algebra. Moreover, using solely the Dolan-Grady relations and the Jacobi identity allows the full infinite-dimensional Lie algebra to be written out explicitly with no further constraints \cite{Davies1,Davies2}. We give this algebra in \eqref{Onsager} below. It is amusing to note that the superintegrable chiral Potts Hamiltonians, the place where this chapter in the story started, are in this language simply
\begin{align}
H_{\rm SI}= Q +\lambda \Qh\ 
\label{HSI}
\end{align}
for some real coupling $\lambda$.

In section 2, we define the self-dual $U(1)$-invariant $n$-state quantum Hamiltonian and show how the Onsager algebra appears as a symmetry algebra. We also demonstrate another remarkable feature connected to the presence of the Onsager algebra: the Hamiltonians can be split into left- and right-moving pieces that commute with each other. These allow the definition of a set of commuting chiral Hamiltonians that interpolate between the ferromagnetic and antiferromagnetic cases while remaining integrable.

In section 3, we start to explore the degeneracies resulting from the Onsager symmetry algebra. Because of the lack of periodicity of the generators, we cannot derive the multiplicities directly. Instead, we show explicitly in the $n=2$ free-fermion case how the degenerate multiplets are $2^N$ dimensional, and present numerical evidence that a similar structure persists for all $n$. 

In section 4, we relate our Hamiltonians to those of the spin-$(n-1)/2$ integrable XXZ chains, and use the correspondence to define a set of commuting transfer matrices. We bring the Onsager algebra into the transfer-matrix setting by showing how transfer matrices built using non-fundamental representations of the quantum-group algebra $U_q(sl_2)$ provide generating functions for the Onsager elements.  

In section 5, we analyse the spectrum using the coordinate Bethe ansatz. In this approach the degeneracies stemming from the Onsager symmetry are a consequence of the appearance of exact $n$-string solutions of the Bethe equations, known \cite{Baxtercompleteness} but not heavily studied. We use these solutions to start understanding how to make precise the structure of the degenerate multiplets.

In section 6, we combine the results of sections 4 and 5 to go further in characterising the degeneracies. In particular, we utilise the $T$-$Q$ relations familiar from integrable models \cite{Baxter82} to define operators that create and annihilate the exact $n$-strings. We then give our conclusions in section 7.


\section{The \texorpdfstring{$U(1)$}{U1}-invariant clock models and their symmetries}
\label{sec:symmetries}
\subsection{The self-dual model}

The Hilbert space for models we study consists of an $n$-state quantum ``spin'' on each of the $L$ sites of a chain, i.e. $(\mathbb{C}^n)^{\otimes L}$. The operators $\tau_j,\sigma_j$ act non-trivially only on the $j^{\text{th}}$ spin, i.e.\ as $\tau_j=1\otimes 1\otimes\dots1\otimes\tau\otimes1\otimes\dots 1$.  They generalise the Pauli matrices and satisfy
\begin{align}
\sigma_j^n=\tau_j^n = 1\,, \qquad 
\sigma_j^\dagger = \sigma_j^{n-1}\,, \qquad 
\tau_j^\dagger = \tau_j^{n-1}\,, \qquad 
\sigma_j \tau_j = \omega \tau_j \sigma_j  \,, \qquad 
\sigma_j \tau_k = \tau_k \sigma_j\,,
\label{algebra}
\end{align}
where the parameter $\omega= e^{2 i \pi /n} $ and $j\ne k$. Very little of what follows will require an explicit matrix representation, but a basis where the $\tau_j$ are all diagonal is given by taking
\begin{align}
\tau= \left(
\begin{array}{cccc}
1 & & & \\
  & \omega  & &\\
    &  & \ddots &\\
      &  & & \omega^{n-1}
\end{array}
 \right)  \,, 
 \qquad 
 \sigma = \left(
\begin{array}{cccc}
0 & 1 & &  \\
  & \ddots  & \ddots  & \\
    &  & \ddots &  1\\
    1  &  & & 0
\end{array}
 \right) \, .
 \label{tausigma}
\end{align}
Thus in this basis $\tau$ can be thought of as measuring the value of the spin, while $\sigma$ shifts it.

The simplest, and most widely studied, version of the $n$-state clock chain has Hamiltonian
\begin{align}
H_{\mathbb{Z}_n} = - \sum_{j=1}^L
 \sum_{a=1}^{n-1} \alpha_a (\tau_j)^a 
 -  \sum_{j=1}^L \sum_{a=1}^{n-1} \widehat{\alpha}_a (\sigma_j^\dagger \sigma_{j+1})^a \,, 
 \label{HZn}
 \end{align}
where hermiticity requires that the couplings obey $\alpha_a^* = \alpha_{n-a}$, $\widehat{\alpha}_a^* = \widehat{\alpha}_{n-a}$. 
This Hamiltonian  is invariant under the global $\mathbb{Z}_n$ symmetry $\tau_j \to \omega \tau_j$.  A famous special case called the $n$-state Potts model arises by equating $\alpha_j=\alpha_{j'}$ and $\widehat{\alpha}_j=\widehat{\alpha}_{j'}$ for all $j,j'$, and so promotes the symmetry to the permutation group $S_n$. However,  \eqref{HZn} need not have any symmetries other than $\mathbb{Z}_n$; for example taking any of the $\alpha_a$ complex breaks time-reversal symmetry, while taking any $\widehat{\alpha}_a$ complex breaks spatial parity symmetry. 

One reason these models were introduced and widely studied is that they generalize the quantum Ising chain (the $n=2$ case) in a fairly natural way.  In particular, they allow for Kramers-Wannier duality \cite{Kramers41}, exchanging high and low temperatures in the corresponding classical model. The most important part of the duality transformation for this translation-invariant system
can be taken to be 
\begin{align}
\tau_j 
\longrightarrow 
\sigma_j^\dagger \sigma_{j+1}\,,\qquad\quad
\sigma_j^\dagger \sigma_{j+1}\longrightarrow 
\tau_{j+1}\ ,
\label{duality}
\end{align} 
up to some subtleties with boundary conditions. The key observation is that the duality transformation preserves
 the algebra \eqref{algebra}. Duality interchanges the two types of terms, and so the model is self-dual with periodic boundary conditions when $\alpha_a  = \widehat{\alpha}_a$ for all $a$.

This Hamiltonian \eqref{HZn} is typically not integrable for $n>2$. A well-known integrable case correspond to the self-dual point of the $n$-state Potts model, where $\alpha_j=\alpha_{j'}=\widehat{\alpha}_j=\widehat{\alpha}_{j'}$. This chain describes the transition (second-order for $n\le 4$ \cite{Baxter82,Duminil15} and first-order for $n>4$ \cite{Duminil16}) between an ordered phase with $S_n$ symmetry breaking and a disordered phase.  A self-dual integrable point with $\mathbb{Z}_n\times \mathbb{Z}_2$ symmetry is critical for all $n$ \cite{Fateev82}, and in the continuum is described by the ``parafermion'' conformal field theory \cite{Fateev85}. There exists a two-parameter integrable deformation \cite{Perk97} called  the ``chiral Potts model'', although the model does not have an $S_n$ symmetry, but in general only $\mathbb{Z}_n$. The superintegrable Hamiltonian \eqref{HSI} is a one-parameter subset of this model.

The purpose of this paper is to analyse in depth another integrable model generalising \eqref{HZn} to have an even larger symmetry, promoting the $\zn$ symmetry to a full $U(1)$ symmetry.  The key observation is that a particular linear combination of the $\tau$ matrix defined in \eqref{tausigma} is a $U(1)$ symmetry generator $S^z$. 
Namely, the operator $Q$ defined by
\begin{align}
Q=\sum_{j=1}^L S^z_j \ ,\qquad\quad S^z_j =\sum_{a=1}^{n-1}\frac{1}{1-\omega^{-a}} (\tau_j)^a 
\label{Qdef}
\end{align}
is a $U(1)$ charge. 
The single-site operator $S^z$ is that occuring in the spin-$(n-1)/2$ representation of the $SU(2)$ algebra, as using the explicit form for $\tau$ in \eqref{tausigma} gives the diagonal $n\times n$ matrix 
whose entries are $(S^z)_{bb'} = \frac{1}{2}(n+1-2b)\delta_{bb'}$.  Acting with $\sigma_j$ on an eigenstate of $Q$ gives states whose eigenvalues of $Q$ either increase by 1 or decreases by $n-1$. Thus the operator $Q$ does not commute with the Hamiltonian \eqref{HZn}, because terms in the latter can violate conservation of $Q$ by $\pm n$.



We instead consider another nearest-neighbour Hamiltonian that does commute with $Q$. The trick is to combine $\tau$ and $\sigma$ operators to remove the $U(1)$-violating processes. The unique such $U(1)$-invariant Hamiltonian with self-duality and only nearest-neighbour interactions is then
\begin{align} 
H_n = 
i \sum_{j=1}^L \sum_{a=1}^{n-1} \frac{1}{1-\omega^{-a}}
 \Bigg[(2a-n) \left(\tau_j^a  + (\sigma_j^\dagger \sigma_{j+1})^a  \right) 
+ \sum_{b=1}^{n-1} \frac{1 - \omega^{-a b}}{1-\omega^{-b}}   \left( \tau_j^b (\sigma_j^\dagger \sigma_{j+1})^a  + (\sigma_j^\dagger \sigma_{j+1})^a \tau_{j+1}^b  \right) \Bigg]\,.
\label{Hnchiral}
\end{align}
It is worth noting that when this Hamiltonian is written in terms of parafermionic operators \cite{Fradkin80,Fendley12}, each term involves at most only three consecutive such operators, explaining  how the model can be both self-dual and nearest-neighbour while still being more complicated than \eqref{HZn}.

While the self-duality of $H_n$ is apparent in the form (\ref{Hnchiral}), the $U(1)$ conservation is not.  Although it is not difficult to show directly that it indeed commutes with $Q$, it is more illuminating to rewrite it in terms of
\begin{align}
 S^+_j\equiv \sigma_j\left(1-\frac{1}{n}\sum_{a=0}^{n-1}(\tau_j)^a\right)\ ,
 \qquad\quad S^-_j= (S^+_j)^\dagger\ .
\end{align}
Acting on a single site in the basis \eqref{tausigma} where $\tau$ is diagonal, $S^\pm$ has matrix elements $(S^\pm)_{bb'}= \delta_{b,b\pm 1}$. These generators therefore satisfy 
\begin{align}
\big[Q,\,S^\pm_j\big] = \pm S^\pm_j\ .
\label{QScomm}
\end{align}
Then we show in the Appendix that $H_n$ can be rewritten in the remarkably simple form
\begin{align}
 H_n = 
 i \sum\limits_{j=1}^L \sum\limits_{a=1}^{n-1} \frac{1}{1-\omega^{-a}} \biggl[ (2a-n) \tau_j^a + n \left(S_j^+S_{j+1}^-\right)^{n-a} - n \left(S_j^- S_{j+1}^+\right)^a \biggr]\,. 
     \label{HnchiralSpm}
\end{align}
Using \eqref{QScomm} shows immediately that the Hamiltonian $H_n$ is $U(1)$ invariant. 
The commutator $[S^+,\,S^-]$ is not proportional to $S^z$, so the three do {not} satisfy the $SU(2)$ commutation relations and the model does not have an $SU(2)$ symmetry. However, we exploit in section \ref{sec:transfermatrices} their connection to representations of the quantum-group algebra $U_q(SL(2))$, a deformation of $SU(2)$.

We refer to the model with Hamiltonian $H_n$ in
\eqref{Hnchiral} or \eqref{HnchiralSpm} as the {\it self-dual $U(1)$-invariant clock model}. This model has appeared before as a particular case of the integrable XXZ chain of spin $(n-1)/2$, as we will detail in section \ref{sec:XXZ}.
For $n=2$, it is bilinear in fermionic operators and so a free theory; we solve it in section \ref{sec:n2}.  For $n=3$, it also has arisen in the study of models based on the Temperley-Lieb algebra \cite{Ikhlef,Ikhlef2}. In a separate paper \cite{Phasediagram}, we will describe the rich physics of a Hamiltonian given by a linear combination of  \eqref{HZn} and  \eqref{HnchiralSpm}.

\subsection{Onsager symmetry}
\label{sec:Onsager}
 
We will devote much of this paper to describing the many interesting symmetries of $H_n$.  One remarkable feature of the Hamiltonian $H_n$ is that despite its being a strongly interacting spin chain for $n>2$, it is quite simple to show that it has a symmetry algebra with an infinite number of generators as $L\to\infty$. This feature arises because $H_n$ is self-dual and commutes with $Q$. It therefore must also commute with the dual of $Q$:
\begin{align} 
\big[\Qh,\,H\big]=0\qquad \hbox{ for }\quad \Qh = \sum_{j=1}^L \sum_{a=1}^{n-1} \frac{1}{1-\omega^{-a}}(\sigma_j^\dagger \sigma_{j+1})^a   \, .
\end{align}
The Hamiltonian thus has a second $U(1)$ symmetry. The interesting symmetries arise because $Q$ and $\widehat{Q}$  do {\em not} commute with each other. Repeatedly commuting $Q$ and $\Qh$ gives rise to an infinite-dimensional Lie algebra called the {\em Onsager algebra} \cite{Onsager44}. 


A remarkable feature of our self-dual Hamiltonian $H_n$ is that since both $Q$ and $\Qh$ commute with it, all the Onsager-algebra elements do as well. Moreover, as explained in the introduction, thinking about $Q$ as generating a $U(1)$ symmetry allows the key relations of the algebra to be found with almost no work. The reason why $Q$ and $\Qh$ do not commute, and why $\Qh$ can be split as (\ref{qhatdecomposition}), is that acting with $(\sigma^\dagger_j\sigma_{j+1})^a$ can change the charge under $Q$ by $\pm n$. This ensuing Dolan-Grady conditions (\ref{Dolan1},\ref{Dolan2}) and the Jacobi identity gives the Onsager algebra, as proved in \cite{Davies2}.

The Onsager algebra is typically given in the form originally found by Onsager. Whereas this is natural if writing the elements in terms of Majorana fermion operators, it obscures the $U(1)$ structure. To make the $U(1)$ structure more apparent, we instead display this algebra in terms of a set of generators $Q^0_m$, $Q^+_m$, and $Q^-_m$, with $m$ an integer and $Q^\pm_{-m} \equiv -Q^{\pm}_m$ and $Q^0_{-m}\equiv Q_m$. Denoting $Q^0_0\equiv 4Q/n$ and $Q^r_1\equiv 4Q^r/n$ for $r=0,\pm$, the Onsager algebra is\footnote{Onsager's convention is to describe the elements by two sets of operators $A_m$ and $G_m=-G_{-m}$. The two generators are $A_0 = 4Q/n$ and $A_1 = 4\Qh/n$, and in general, our elements are related by $Q^0_m=(A_m+A_{-m})/2$ and $Q^\pm_m = (A_m-A_{-m} \pm 2G_m)/4$.} 
\begin{align}
[{Q}_l^r, {Q}_m^r] &= 0   \cr
\big[{Q}_l^-, {Q}_m^+\big] &= {Q}_{m+l}^0 -  {Q}_{m-l}^0   \cr
\big[{Q}_l^-,{Q}_m^0\big] &=2 \Big( {Q}_{m+l}^- -  {Q}_{m-l}^-  \Big)     \cr
\big[{Q}_l^+, {Q}_m^0\big] &= 2 \Big( {Q}_{m-l}^+ -  {Q}_{m+l}^+  \Big)   \label{Onsager} \ .
\end{align}

We are not aware of a closed-form expression of the $Q^r_m$ in the clock models. However, like the Hamiltonian, the $Q^r$ have a nice expression in terms of $S^\pm_j$:
\begin{align}
Q^0 &=   \sum_{j=1}^L  \sum_{a=1}^{n-1} 
 \frac{1}{1-\omega^{-a}} 
\Big[ 
( S_j^- S_{j+1}^+)^{a}  - \omega^{-a}( S_j^+ S_{j+1}^-)^{a}
\Big] 
\ ,\cr
Q^+ &=   \sum_{j=1}^L  \sum_{a=1}^{n-1} 
\frac{1}{1-\omega^a } 
 (S_j^+)^{a} (S_{j+1}^+)^{n-a}  
 \ ,
\label{QSpm}
\end{align}
with $Q^-=(Q^+)^\dagger$ .

It should be expected that such a rich, non-abelian symmetry of our models should come with interesting physical consequences. We will start to examine these in section \ref{sec:spectrum}.




\subsection{Chiral decomposition}
\label{sec:chiraldecomposition}

Comparing the explicit expressions (\ref{HnchiralSpm}) for $H_n$ and (\ref{QSpm}) for $Q^0$ leads to another interesting feature of the model:  the $H_n$ can be split into two commuting pieces.  Namely, 
define
\begin{align}
 H_{\rm R} = 
 i \sum_{j=1}^L  \sum_{a=1}^{n-1} 
\frac{1}{1-\omega^{-a}} 
\Big[
 n  
 \left(  S_{j}^- S_{j+1}^+\right)^a 
  +  
    \frac{1}{2}(2a-n)
 \left(    \tau_j\right)^a 
     \Big] \ ,\qquad\quad H_{\rm L}=(H_{\rm R})^\dagger\ .
     \label{HRdef}
 \end{align}    
We used the subscripts $\rm R$ and $\rm L$ because the non-diagonal pieces in
$H_{\rm  L}$ and $H_{\rm  R}$ contain the parts of $H_n$ that carry $U(1)$ charge toward the right and the left respectively. These operators were chosen so that
\begin{align}
H_n &= H_{\rm  R} +  H_{\rm L} \,,\\
Q^0 &= \frac{i}{n} \left(H_{\rm R}  - H_{\rm L}  \right)  \,,
\label{decomposition} 
\end{align}
As described in section \ref{sec:Onsager}, $[Q^0,\,H_n]=0$ by construction. Thus we immediately find 
\be 
[ H_{\rm  R} ,  H_{\rm L} ]    =  0  \,. 
\label{HLHR0}
\ee
In other terms, the Hamiltonians $H_n$ can be split as the sum of a left- and right-moving parts that commute with each other!  It is worth noting that while this decomposition holds for twisted boundary conditions as well as periodic, the analogous $H_{\rm L}$ and $H_{\rm R}$ do not commute for open boundary conditions. 
 
Thus defining
\be 
H(\alpha) =  e^{i \alpha} H_{\rm R} + e^{-i \alpha} H_{\rm L} \,,
\label{Halpha}
\ee 
gives a one-parameter family of commuting Hamiltonians obeying $H(0)=H_n$ and $H(\pi)=-H_n$, while the ``maximally chiral''  Hamiltonian $H(\pi/2)$  is proportional to $Q^0$. Since these Hamiltonians all commute with one another, they share the same eigenspaces, a fact that will prove quite useful in our analysis.
However, it is important to note that $H(\alpha)$ commutes with all the Onsager generators only for $\alpha=0$ or $\pi$; only the $Q^0_m$ commute with $H(\alpha)$ for all $\alpha$.

Another decomposition of the Hamiltonian as the sum of two commuting pieces has been found for the case $n=3$, in terms of Temperley-Lieb generators \cite{Ikhlef}. Interestingly, this splitting is different from the one presented here, or from any linear combination of $H_{\rm R}$ and $H_{\rm L}$. The two commuting Hamiltonians presented in \cite{Ikhlef} do not conserve the $U(1)$ charge individually, and so might signal an additional symmetry of our models. 

\subsection{The Onsager algebra for \texorpdfstring{$n=2$}{n=2}}

To give a little more intuition into the Onsager algebra, we write its generators out explicitly in the $n=2$ case using fermionic operators.
The $U(1)$-invariant self-dual Hamiltonian for $n=2$ in terms of Pauli matrices is
\begin{align}
H_2 = \frac{1}{2}\sum_{j=1}^L\left(\sigma_j^x\sigma_{j+1}^y-\sigma_j^y\sigma_{j+1}^x\right)=i\sum_{j=1}^L\left(\sigma_j^+\sigma_{j+1}^--\sigma_j^-\sigma_{j+1}^+\right).
\end{align}
The $U(1)$ charge operators commuting with $H_2$ are
\begin{align}
Q=\frac{1}{2}\sum_{j=1}^L\sigma_j^z\ ,\qquad\quad \Qh=\frac{1}{2}\sum_{j=1}^L\sigma_j^x\sigma^x_{j+1}\ .
\end{align}
This Hamiltonian can be split into two as $H_2=H_{\rm L}+H_{\rm R}$, where
\begin{align}
H_{\rm L}=i\sum_{j=1}^L\sigma_j^+\sigma_{j+1}^-\,,\qquad H_{\rm R}=-i\sum_{j=1}^L\sigma_j^-\sigma_{j+1}^+=(H_{\rm L})^\dagger\,.
\label{split2}
\end{align}
It is simple to check that $[H_{\rm L},H_{\rm R}]=0$ for periodic boundary conditions (but not for open).

To give a nice expression for the Onsager elements, we use a Jordan-Wigner transformation to complex fermions:
\begin{align}
c_j=\sigma_j^-\prod\limits_{l<j}\sigma_l^z\ , \qquad c_j^\dagger=\sigma_j^+\prod\limits_{l<j}\sigma_l^z\ .
\end{align}
These operators obey the usual anticommutation relations 
\be
\{c_j,c_l\}=\{c_j^{\dag},c_l^{\dag}\}=0\,, \qquad \{c_j,c_l^{\dag}\}=\delta_{jl}.
\ee
In terms of the fermions, the $U(1)$ charge is simply 
\begin{align}
Q   
=-\frac{L}{2}+\sum_{j=1}^L c^\dagger_j c_j\ ,
\end{align}
so $Q$ up to a shift measures the fermion number. The commutator of the fermions with $Q$ is simple, namely $[Q,c_j^\dagger]=c^\dagger_j$ and $[Q,c_j]=-c_j$. For simplicity we assume that $L$ is even, so that the eigenvalues of $Q$ are integers. 
The Hamiltonian is then 
\begin{align}
H_2=i\sum_{j=1}^{L-1} \left(c^\dagger_j c^{}_{j+1} + c_j^{}c^\dagger_{j+1}\right)-i(-1)^Q\left(c^\dagger_L c^{}_{1} + c^{}_Lc^\dagger_{1}\right)
\label{H2ferm}
\end{align}
where $(-1)^Q=\prod_j \sigma_j^z$ in this basis measures whether the number of spin-down particles is even or odd, or, equivalently, fermion-number parity. This twist factor $-(-1)^Q$ arises because of the non-locality of the map from spins to fermions.

In terms of the fermions, the dual $U(1)$ charge is
\begin{align}
\Qh=\frac{1}{2}\sum\limits_{j=1}^{L}(-1)^{{\cal T}_{j+1}}\left(c_j-c_j^{\dag}\right)\left(c_{j+1}+c_{j+1}^{\dag}\right) .
\end{align}
where the twisting is defined by
\[{\cal T}_{s}=(Q+1)\lfloor(s-1)/L \rfloor
\]
with $\lfloor x \rfloor$ the floor of $x$. In this form it is obvious how to split $\Qh$ into the $Q^r$: all terms involving any $c_j^\dagger c_{j+1}^\dagger$ are contained in $Q^+$, all with $c_jc_{j+1}$ are in $Q^-$, with the others having zero charge and so in $Q^0$.
Since commutators of bilinears in fermions give bilinears, the Onsager elements are also bilinears in the fermions. A little bit of algebra then yields
\begin{align}
Q_m^0=(-1)^m\sum\limits_{j=1}^L(-1)^{{\cal T}_{j+m}}\left(c_j^{\dag}c_{j+m}-c_jc_{j+m}^{\dag}\right)\,,
\qquad Q_m^+=(-1)^m\sum\limits_{j=1}^L(-1)^{{\cal T}_{j+m}} c_j^{\dag}c_{j+m}^{\dag}\,,\quad 
\label{Qmfermion}
\end{align}
where $Q_m^-=(Q_m^+)^\dagger$ as always. It is thus obvious that $[Q,Q_m^r]= 2 rQ_m^r$. 

From these explicit expressions it is also clear that the Onsager elements are periodic: $Q^r_{m+L}=-(-1)^Q Q^r_m$, and so $Q^r_{m+2L}=Q_m^r$. Intuitively, one can think each shift by $L$ as wrapping the Jordan-Wigner string around one more time. Such elegant periodicity in $m$ is a consequence of the free-fermion nature of $n=2$; we have verified by brute force that for general $n$ there is no such linear relation among Onsager elements under shifts linear in $L$.

\comment{
In the presence of the $U(1)$ symmetry, it is natural to use Dirac fermions
\be
c_j=\frac{1}{2}\left(\gamma_{2j-1}-i\gamma_{2j}\right)\,,\qquad c_j^{\dag}=\frac{1}{2}\left(\gamma_{2j-1}+i\gamma_{2j}\right)\, 
\ee
\[\gamma_{2j-1}=\sigma_j^x\prod\limits_{l<j}\sigma_l^z\ , \qquad \gamma_{2j}=i\sigma_j^x\sigma_j^z\prod\limits_{l<j}\sigma_l^z\ .\]

$\{\gamma_a,\gamma_b\}=2\delta_{jk}$. The Hamiltonian becomes
\begin{align}
H_2=-\frac{i}{2}\sum_{j=1}^{2L-2}\gamma_j\gamma_{j+2} +\frac{i}{2}(-1)^Q\left(\gamma_{2L-1}\gamma_1+\gamma_{2L}\gamma_2\right),

\end{align}
It is worth noting that (\ref{H2def}) makes apparent another way of splitting $H_2$ into two
commuting Hamiltonians, simply by considering terms with even and odd $j$. The resulting Hamiltonians however are individually Hermitian, and so not equivalent to those in \eqref{split2}. 
}

\section{The degeneracies}
\label{sec:spectrum}

The many symmetries described in section \ref{sec:symmetries} suggest that the self-dual Hamiltonians $H_n$ are integrable. In section \ref{sec:betheansatz} we use the Bethe ansatz to show that indeed this is so. 
Moreover, the fact that the symmetry generators obey a non-abelian algebra indicates that there should be degeneracies in the spectrum. Namely, since the Onsager generators $Q_m^\pm$ do not commute with $Q$, acting with them on an energy eigenstate must give another state with the same energy but with charge changed by $\pm n$.  The purpose of this section is to characterise these degeneracies in a simple manner, before plunging into the detailed technical analysis.

\subsection{The degeneracies for \texorpdfstring{$n=2$}{n=2}}
\label{sec:n2}

It is highly illuminating to start by analysing the $n=2$ case. Since the Hamiltonian $H_2$ in (\ref{H2ferm}) is bilinear in free-fermion operators, the entire spectrum can be computed, and the degeneracies due to the Onsager algebra can be isolated.

The Hamiltonian can be diagonalised by Fourier transforming these fermions as
\begin{align}
c_k=\frac{1}{\sqrt{L}}\sum\limits_ke^{ij\left(k-\frac{\pi}{2}\right)}c_j\,,\qquad c_k^{\dag}=\frac{1}{\sqrt{L}}\sum\limits_ke^{-ij\left(k-\frac{\pi}{2}\right)}c_j^{\dag}\,,
\end{align}
with $k=2m\pi/L+\pi/2$ for $(-1)^Q=-1$ and $k=(2m+1)\pi/L+\pi/2$ for $(-1)^Q=1$ and we have added the extra $\pi/2$ for later convenience. We then find
\begin{align}
H_{\rm L}=i\sum\limits_ke^{-i\left(k-\frac{\pi}{2}\right)}n_k\,,\qquad H_{\rm R}=-i\sum\limits_ke^{i\left(k-\frac{\pi}{2}\right)}n_k,
\end{align}
where $n_k=c_k^{\dag}c_k$ is the fermion number operator. Thus
\begin{align}
H=-2\sum\limits_k n_k\cos k\ ,\qquad\quad Q=-\frac{L}{2} +\sum_k n_k\ .
\end{align}
To ensure the correct boundary conditions, $k=2m\pi/L$ for spin parity $(-1)^Q=-1$, while $k=(2m+1)\pi/L$ for $(-1)^Q=1$. It should be noted that particles of energies $\pm k$ have equal energies, while those of $k$ and $\pi-k$ have opposite energies, a fact which will be of crucial importance. Acting on a state with $c^\dagger_k$ or $c_k$ will therefore either annihilate the state or change the charge, that is the total fermion number, by $\pm 1$, respectively.

The ground states in each spin-parity sector are then found by filling all of the negative-energy levels ($\left|k\right|<\pi/2$), while leaving the positive ones empty. There are a few subtleties here on  ground-state degeneracies arising from zero modes, but these are unimportant for the subsequent discussion. To obtain excited states in each sector, we then act with pairs of annihilation and creation operators: $c_kc_q$, $c_kc_q^{\dag}$ or $c_k^{\dag}c_q^{\dag}$. This allows us to generate the full spectrum of the model.

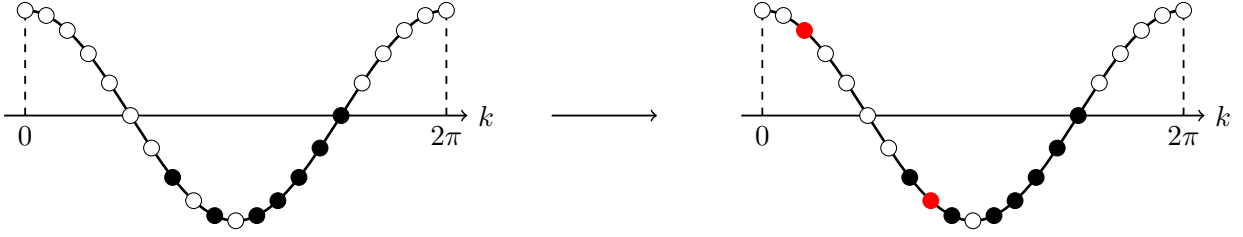
\begin{figure}[tb]
\begin{center}
\begin{tikzpicture}[scale=1.4]
\draw[black, line width=1] (0,1) cos (1,0);
\draw[black, line width=1] (1,0) sin (2,-1);
\draw[black, line width=1] (2,-1) cos (3,0);
\draw[black, line width=1] (3,0) sin (4,1);
\draw[->,black, line width=0.7] (-0.2,0)-- (4.2,0) node[right] {$k$};
\draw[dashed,black, line width=0.7]  (0,0)node[below]  {$0$} -- (0,1);
\draw[dashed,black, line width=0.7]  (4,0)node[below]  {$2\pi$} -- (4,1);
\draw[fill=white] (0,1) circle (0.075);
\draw[fill=white] (0.2,0.951) circle (0.075);
\draw[fill=white] (0.4,0.809) circle (0.075);
\draw[fill=white] (0.6,0.5878) circle (0.075);
\draw[fill=white] (0.8,0.309) circle (0.075);
\draw[fill=white] (1,0) circle (0.075);
\draw[fill=white] (1.2,-0.309) circle (0.075);
\draw[fill=black] (1.4,-0.5878) circle (0.075);
\draw[fill=white] (1.6,-0.809) circle (0.075);
\draw[fill=black] (1.8,-0.951) circle (0.075);
\draw[fill=white] (2,-1) circle (0.075);
\draw[fill=black] (2.2,-0.951) circle (0.075);
\draw[fill=black] (2.4,-0.809) circle (0.075);
\draw[fill=black] (2.6,-0.5878) circle (0.075);
\draw[fill=black] (2.8,-0.309) circle (0.075);
\draw[fill=black] (3,0) circle (0.075);
\draw[fill=white] (3.2,0.309) circle (0.075);
\draw[fill=white] (3.4,0.5878) circle (0.075);
\draw[fill=white] (3.6,0.809) circle (0.075);
\draw[fill=white] (3.8,0.951) circle (0.075);
\draw[fill=white] (4,1) circle (0.075);

\draw[->, line width=0.7] (5,0) -- (6,0);

\begin{scope}[shift={(7,0)}]
\draw[black, line width=1] (0,1) cos (1,0);
\draw[black, line width=1] (1,0) sin (2,-1);
\draw[black, line width=1] (2,-1) cos (3,0);
\draw[black, line width=1] (3,0) sin (4,1);
\draw[->,black, line width=0.7] (-0.2,0)-- (4.2,0) node[right] {$k$};
\draw[dashed,black, line width=0.7]  (0,0)node[below]  {$0$} -- (0,1);
\draw[dashed,black, line width=0.7]  (4,0)node[below]  {$2\pi$} -- (4,1);

\draw[fill=white] (0,1) circle (0.075);
\draw[fill=white] (0.2,0.951) circle (0.075);
\draw[red,fill=red] (0.4,0.809) circle (0.075);
\draw[fill=white] (0.6,0.5878) circle (0.075);
\draw[fill=white] (0.8,0.309) circle (0.075);
\draw[fill=white] (1,0) circle (0.075);
\draw[fill=white] (1.2,-0.309) circle (0.075);
\draw[fill=black] (1.4,-0.5878) circle (0.075);
\draw[red,fill=red] (1.6,-0.809) circle (0.075);
\draw[fill=black] (1.8,-0.951) circle (0.075);
\draw[fill=white] (2,-1) circle (0.075);
\draw[fill=black] (2.2,-0.951) circle (0.075);
\draw[fill=black] (2.4,-0.809) circle (0.075);
\draw[fill=black] (2.6,-0.5878) circle (0.075);
\draw[fill=black] (2.8,-0.309) circle (0.075);
\draw[fill=black] (3,0) circle (0.075);
\draw[fill=white] (3.2,0.309) circle (0.075);
\draw[fill=white] (3.4,0.5878) circle (0.075);
\draw[fill=white] (3.6,0.809) circle (0.075);
\draw[fill=white] (3.8,0.951) circle (0.075);
\draw[fill=white] (4,1) circle (0.075);
\end{scope}
\end{tikzpicture}
\end{center}
\caption{
Action of the operator $\mathcal{Q}^{\dag}(k)$ on a given eigenstate of the $n=2$ Hamiltonian. If the levels $k,\pi-k$ were vacant in the original state, $\mathcal{Q}(k)$ creates a degenerate eigenstate with two more fermions (in red), increasing the $U(1)$ charge by 2. 
}
\label{fig:fermions}
\end{figure}

Understanding the degeneracies due to the Onsager algebra is straightforward in terms of the fermions.
Because of the periodicity of the Onsager algebra $Q^r_{m+2L}=Q_m^r$ for $n=2$, the 
momentum-space versions of the Onsager elements are quite simple. Using the explicit expressions \eqref{Qmfermion} and defining
\begin{align}
\mathcal{Q}(k)\equiv-\frac{i}{\sqrt{L}}\sum\limits_{m=1}^{L-1}\sin \biggl(m\left(k+\frac{\pi}{2}\right)\biggr)Q_m^{-}
\label{SQ}
\end{align}
gives
\begin{align}
\mathcal{Q}^{\dag}(k)=c_k^{\dag}c_{\pi-k}^{\dag}\ ,\qquad\quad \mathcal{Q}(k)=c_{\pi-k}c_k\ .
\label{Sdef}
\end{align}
Acting with these operators 
leaves the energy invariant, changes the momentum by $\pi$, and alters the $U(1)$ charge by $\pm 2$. 
An example of the action of $\mathcal{Q}^\dagger$ on a particular state is illustrated in Figure \ref{fig:fermions}.  This action is non-trivial only on states with both a hole in the Fermi sea at momentum $k$, and no particle at momentum $\pi-k$. Acting with $\mathcal{Q}(k)$ is non-trivial only on states with a filled level at momentum $\pi-k$ and without a hole at $k$. In terms of the usual quasiparticle picture, $\mathcal{Q}^\dagger(k)$ creates a particle and annihilates an antiparticle, and vice versa for $\mathcal{Q}(k)$. Such an action indeed changes the charge by $+2$ and $-2$ respectively, while leaving the energy invariant.


Applying the $\mathcal{Q}^{\dag}(k)$ and $\mathcal{Q}(k)$ operators to general states clearly leads to degeneracies. Since $\mathcal{Q}(\pi-k)=-\mathcal{Q}(k)$, we can restrict consideration to $|k|<\pi/2$. 
To find the full structures of the multiplets with these degeneracies, consider an energy eigenstate $|s_{\rm min}\rangle$ annihilated by all $\mathcal{Q}(k)$. In this state, each pair of levels $k$ and $\pi-k$ can be occupied by at most one fermion. Let $N_s$ be the number of such pairs completely unoccupied, and $N'_s$ the number of pairs with at exactly one level occupied. Since there are $L$ levels, $N_s+N_s'=L/2$ with our assumption that $L$ is even.
The charge of this state must therefore be
\[Q_{\text{min}}=-L/2+N_s' = -N_s\ .\] 
There are $N_s$ different values of $k$ such that $\mathcal{Q}^\dagger(k)|s_{\rm min}\rangle\ne 0$. Acting with any of these once increases the charge by $2$, giving a multiplet of states with the same energy. Since  $[\mathcal{Q}^\dagger(k),\mathcal{Q}^\dagger(k')]=0$ and $(\mathcal{Q}^\dagger(k))^2=0$, the total degeneracy of this multiplet is $2^{N_s}$, with the number of states $d_p$ at each charge $Q=2p-N_s$ given by
\be 
d_p = {{N_s}\choose{p}} \ .
\ee


The relation (\ref{SQ},\ref{Sdef}) between the Onsager elements and fermion bilinears is nice because of the periodicity of the Onsager elements under $m\to m+2L$, a property that does not generalise to arbitrary $n$. Nonetheless, degeneracies analogous to these for $n=2$ are not simply a free-fermionic fluke, and the subject of the rest of the paper.

\subsection{Structure of degeneracies for general \texorpdfstring{$n$}{n}}
\label{sec:degeneracies}

Degeneracies should be expected as a general feature of models with a non-abelian symmetry, but constructing the analog of $\mathcal{Q}^\dagger(k)$ for $n>2$ requires considerable work. In sections \ref{sec:betheansatz} and \ref{sec:stringcreation} we give this construction by utilising exact $n$-string solutions of the Bethe equations. Happily, the detailed calculation is not needed to understand the degeneracies qualitatively. Thus we start our general analysis by giving here some numerics for $n=3$ that illustrate this structure nicely.

The Onsager elements $Q^{\pm}_m$ shift the charge by $\pm n$, but still commute with the Hamiltonian. Thus we expect degenerate multiplets with charges differing by multiples of $n$. Some numerical results for $H_3$ using exact diagonalisation can be found in tables \ref{table1} and \ref{table2}. The presence of such degeneracies is readily apparent in both of these. 
\begin{table}[h]
\resizebox{\columnwidth}{!}{%
\begin{tabular}{|c|c|c|c|c|c|c|}
\hline
$Q=-6$ & $Q=-3$ & $Q=0$ & $Q=3$ & $Q=6$ & $L\to\infty$ & CFT\\
\hline
&  & 0 & & & 0 & 0 \\
\hline
& & 0.992453634448 & & & 1.0014 & 1\\
\hline
& & 1.979146217630 & & & 2.0040 & 2\\
\hline
2.870426956543& 2.870426956543 $\times 4$ & 2.870426956543 $\times 6$& 2.870426956543  $\times 4$& 2.870426956543 & 3.0309 & 3 \\
\hline
\end{tabular}
}

\begin{center}
\resizebox{.7\columnwidth}{!}{%
\begin{tabular}{|c|c|c|c|c|}
\hline
$Q=-2$ & $Q=1$ & $Q=4$ & $L\to\infty$ & CFT\\
\hline
0.334282995064 & & & 0.3334 & 1/3 \\
\hline
1.313397175669 & 1.313397175669 $\times 2$ & 1.313397175669 & 1.3364 & 4/3\\
\hline
2.2629252374614 & 2.2629252374615 $\times 2$ & 2.2629252374616 & 2.3436 & 7/3\\
\hline
\end{tabular}
}
\end{center}
\caption{Low-lying energy levels of $H_3$ for $L=16$ with momentum $k=0$ in sectors of various $Q$; the spectra for $Q\to-Q$ are identical. The energies are shifted so the ground-state energy is 0 and rescaled by $L/(2\pi v_F)$, where $v_F=9/2$ is the Fermi velocity. The $\times m$ indicates that there are $m$ levels with this energy, up to differences $< 10^{-10}$. The $L\to \infty$ column gives the extrapolation of the energy to infinite lattice length from a quadratic fit in $1/L$ of the values for $L=12,14,16$. The last column consists of the predictions from the c=3/2 CFT.}
\label{table1}
\end{table}

To make the results even more informative, we have shifted all the energies by a constant (the same in all sectors), and rescaled them by $L/(2\pi v_F)$, where $v_F$ is a ``Fermi'' velocity $v_F=9/2$. This value for the velocity turns out to be derivable using the Bethe ansatz, but here can be simply viewed as a rescaling that reveals a striking feature beyond the degeneracies: levels within a sector are typically approximately split by integers (or half-integers in a few cases). 
This splitting by integers leads to the expectation that the continuum limit of this spin chain is described by a conformal field theory (CFT). This limit is also implied by the mapping on to the spin-$(n-1)/2$ XXZ chain described in section \ref{sec:XXZ}. Earlier work indicates that $H_n$ should scale to a CFT Hamiltonian with central charge $3(n-1)/(n+1)$, while $-H_n$ scales to one with central charge $1$ \cite{XXZSCFT,XXZSCFTFrahm}. In both cases, we have refined and checked these predictions, identifying exactly which CFT it is (including finding the radius of the bosonic field present). We thus include in the tables a column which gives the energies for this level in the corresponding CFT, and defer further analysis of the CFTs to future work.  Worth noting, however, is that the CFT degeneracies are even larger than those on the lattice, as indicated in Table  \ref{table2}, where levels distinct on the lattice but presumably degenerate in the CFT are separated by dashed horizontal lines. 

\begin{table}[h]
\resizebox{\columnwidth}{!}{%
\begin{tabular}{|c|c|c|c|c|}
\hline
$Q=-3$ & $Q=0$ & $Q=3$ & $L\to\infty$ & CFT \\
\hline
0.89191865865 & 0.89191865865 $\times 2$ & 0.89191865865 & 0.8751 & 7/8 \\
\hline
2.67243586227 $\times 2$ & 2.67243586227 $\times 4$ & 2.67243586227 $\times 2$ & 2.9072 &  \\
\cdashline{1-4}
2.77803728576 & 2.77803728576 $\times 2$ & 2.77803728576 & 2.8865 & 23/8 \\
 \cdashline{1-4}
2.86395394892 & 2.86395394892 $\times 2$& 2.86395394892 & 2.8863 & \\
\hline
\end{tabular}
\quad
\begin{tabular}{|c|c|c|c|c|}
\hline
$Q=-5$ & $Q=-2$ & $Q=1$ & $L\to\infty$ & CFT \\
\hline
&  & 0.211743760 & 0.2084 & $\frac{5}{24}\approx 0.2083$ \\
\hline
& 2.0994556102 $\times 2$ & 2.0994556102 $\times 2$ & 2.2202 & \\
\cdashline{1-4}
2.2088764136 & 2.2088764136 $\times 2$ & 2.208876413 & 2.2121 & $\frac{53}{24}\approx 2.2083$\\
\cdashline{1-4}
& & 2.232900511518 & 2.2133 & \\
\hline
\end{tabular}
}
\caption{Low-lying energy levels of $H_3$ as in table \ref{table1}, except with $k=\pi$.}
\label{table2}
\end{table}

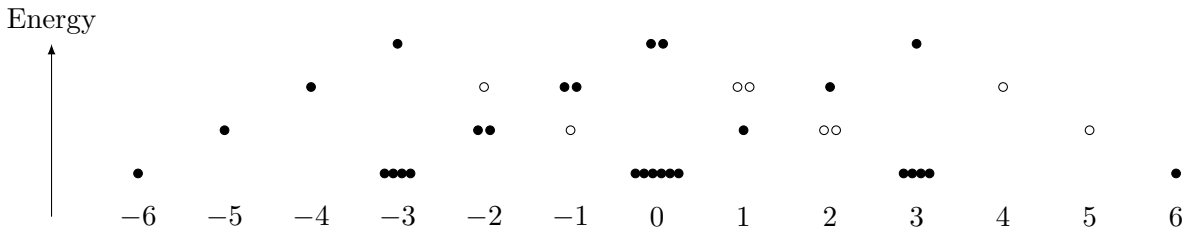
\begin{figure}[ht]
\begin{center}
\begin{tikzpicture}[scale=1.15]
\draw[fill=black,-latex] (-7,-1) -- (-7,1) node[above] {Energy};
\node at (-6,-1) {$-6$};
\node at (-5,-1) {$-5$};
\node at (-4,-1) {$-4$};
\node at (-3,-1) {$-3$};
\node at (-2,-1) {$-2$};
\node at (-1,-1) {$-1$};
\node at (0,-1) {$0$};
\node at (1,-1) {$1$};
\node at (2,-1) {$2$};
\node at (3,-1) {$3$};
\node at (4,-1) {$4$};
\node at (5,-1) {$5$};
\node at (6,-1) {$6$};
\foreach \x in {-0.25,-0.15,-0.05,0.05,0.15,0.25}
{ \draw[fill=black] (\x,-0.5) circle (0.05);
}
\foreach \x in {-0.15,-0.05,0.05,0.15}
{ \draw[fill=black] (-3+\x,-0.5) circle (0.05);
\draw[fill=black] (3+\x,-0.5) circle (0.05);
}
\foreach \x in {0}
{ \draw[fill=black] (-6+\x,-0.5) circle (0.05);
\draw[fill=black] (6+\x,-0.5) circle (0.05);
}

\draw[black,fill=black] (-5,-0.) circle (0.05);
\draw[black,fill=black] (-2.07,-0.) circle (0.05);
\draw[black,fill=black] (-1.93,-0.) circle (0.05);
\draw[black,fill=black] (1,-0.) circle (0.05);

\draw[black] (5,-0.) circle (0.05);
\draw[black] (2.07,-0.) circle (0.05);
\draw[black] (1.93,-0.) circle (0.05);
\draw[black] (-1,-0.) circle (0.05);

\draw[black,fill=black] (-4,0.5) circle (0.05);
\draw[black,fill=black] (-1.07,0.5) circle (0.05);
\draw[black,fill=black] (-0.93,0.5) circle (0.05);
\draw[black,fill=black] (2,0.5) circle (0.05);

\draw[black] (4,0.5) circle (0.05);
\draw[black] (1.07,0.5) circle (0.05);
\draw[black] (0.93,0.5) circle (0.05);
\draw[black] (-2,0.5) circle (0.05);

\draw[fill=black] (-3,1) circle (0.05);
\draw[fill=black] (-0.07,1) circle (0.05);
\draw[fill=black] (0.07,1) circle (0.05);
\draw[fill=black] (3,1) circle (0.05);

\end{tikzpicture}
\end{center}
\caption{
Schematic representation of the degeneracies for the $n=3$ model on a chain of $L=6$ sites. The numbers at the bottom represent the charge $Q$, and full and empty circles are used to distinguish between different multiplets at the same energy. Degeneracies occur between sectors of charge differing by multiples of $n$, and the number of states of a given degenerate tower in each sector is given by a binomial coefficient.}
\label{fig:degeneracies}
\end{figure}

We have presented numerical data for $H_3$, but we have checked $-H_3$ as well as higher $n$. We find that in all cases, degeneracies occur between states of $U(1)$-charge differing by multiples of $n$, exact up to high numerical precision. All states can be grouped into degenerate multiplets characterised by a ``highest-weight'' and ``lowest-weight" pair in the sectors of charge $Q_{\rm max}$ and $Q_{\rm min}$ respectively, such that $Q_{\rm max}-Q_{\rm min}=Nn$, for some integer $N$. We find that for a given degenerate multiplet, the number of states inside the sector with $Q-Q_{\rm min} = p n$ is given by 
\be 
d_p = {{N}\choose{p}} \,. 
\label{binomials}
\ee  
The total degeneracy of the tower is therefore 
\be 
d = \sum_{p=0}^N {{N}\choose{p}}   = 2^N \,
\label{binomials2}
\ee
just as in the free-fermion case of $H_2$.
The structure of multiplets is illustrated schematically in Figure \ref{fig:degeneracies}. For $N$ even, there is a unique $Q_c$ with maximal $d_Q$, while for $N$ odd there are two values $Q_{c_1}$ and $Q_{c_2}$ with maximal $d_Q$. The latter values are the ``centre(s)" of the multiplet and are always found to have values $-n<Q_c<n$, $(-n<Q_{c_1}<Q_{c_2}<n)$.
If $Q_{\rm max}$ is not a multiple of $n$, neither is $Q_{\rm min}$ and there is a second multiplet degenerate with the first but with all $Q\to -Q$.


The multiplicities behave in essentially the same fashion for all $n$. However, the models do not have a free-fermionic interpretation for $n>2$, and instead are strongly interacting, as will become apparent via  the Bethe-ansatz analysis of these models in section \ref{sec:betheansatz}. The underlying reason for this structure seems to have little to do with fermions, free or not, and everything to do with Onsager. Indeed the Onsager algebra \eqref{Onsager} is independent of $n$, so its allowed representations will be independent as well. Free fermions give irreducible representations of the algebra of dimension $2^N$, and we know of no other representations. Thus it should not be surprising that for any $n$ the only representations that appear are of dimension $2^N$.

\subsection{Splitting the degeneracies}
\label{sec:chiralnumerics}

As discussed in section \ref{sec:chiraldecomposition}, the Hamiltonian $H_n$ can be split into two commuting chiral pieces, so that a one-parameter family $H(\alpha)$ of Hamiltonians can be constructed. 
Although the charge-neutral Onsager elements $Q^0_m$ still commute with $H(\alpha)$, the charged elements ${Q_m^\pm}$ do not. Since the latter are what give the exact lattice degeneracies, we expect that these degeneracies are split when $\alpha\ne 0,\pi$. This splitting turns out to be a valuable tool in gaining further insight into these degeneracies.

\begin{figure}[t]
\begin{center}
\includegraphics[scale=0.7]{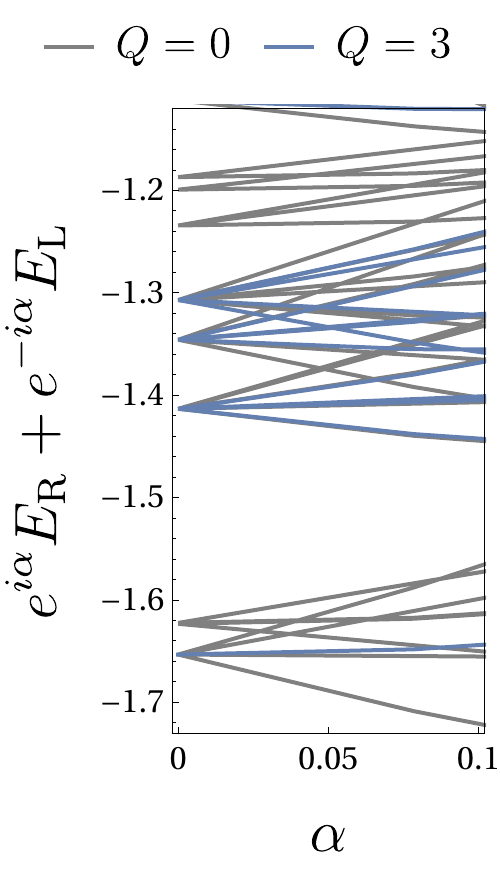}
\hspace{1cm}
\includegraphics[scale=0.4]{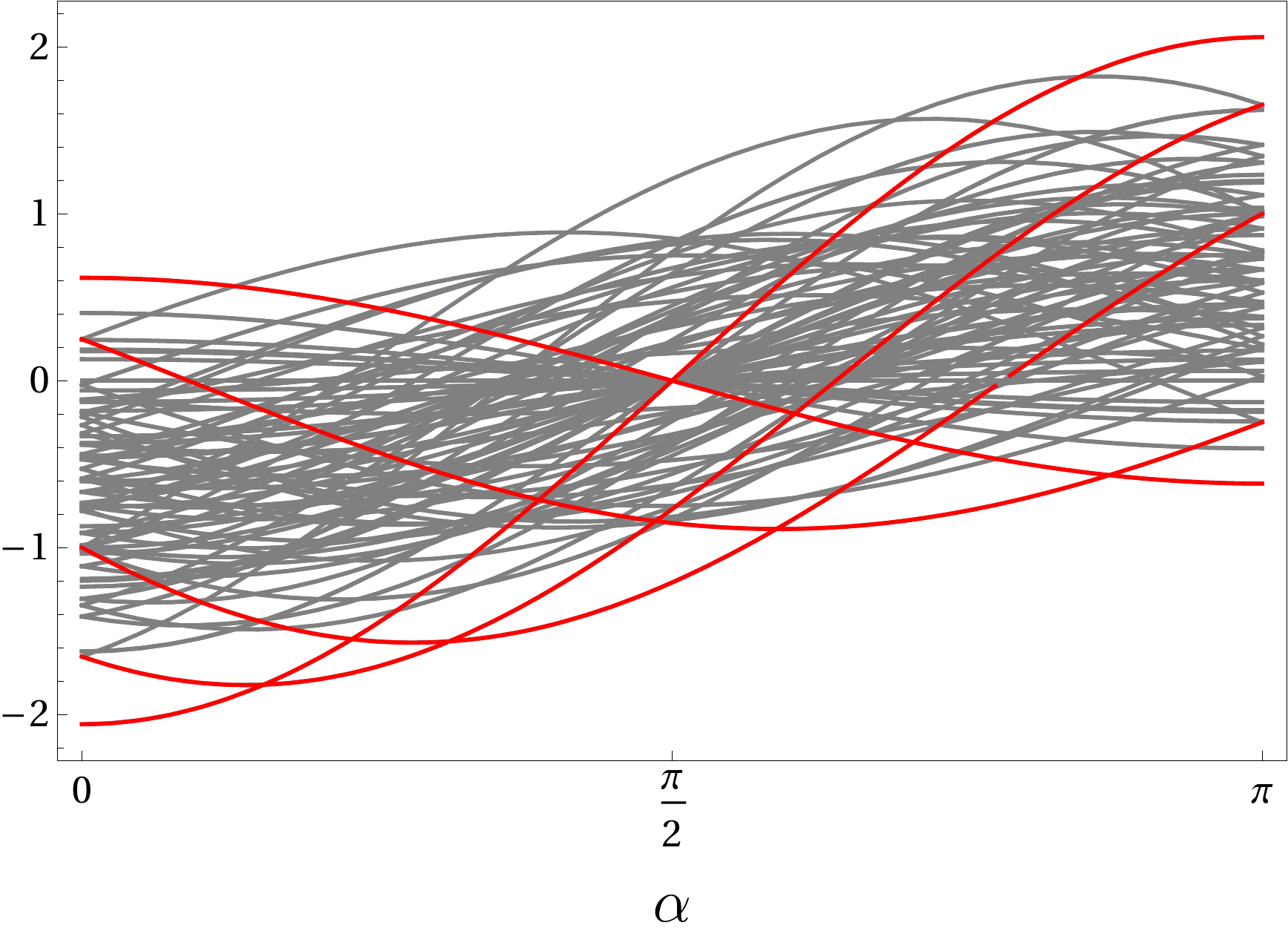}
\end{center}
\caption{
Complete spectrum of the Hamiltonian $H(\alpha))$ for $n=3$ and $L=6$ sites.
The left plot shows the lifting of degeneracies between various sectors when $\alpha$ increases from $0$. 
The right plot shows the levels of the $Q=0$ sector as $\alpha$ is varied between $0$ and $\pi$. The successive ground states are highlighted in red. 
}
\label{fig:crossings0}
\end{figure}

To give an illustration, we plot the spectrum at $L=6$ as a function of $\alpha$ is represented on Figure \ref{fig:crossings0}.  From the left panel, we observe that the degeneracies are indeed lifted for $\alpha \neq 0$. The evolution of the spectrum of \eqref{Halpha} as $\alpha$ is varied between $0$ and $\pi$ is illustrated on the right panel of Figure \ref{fig:crossings0}, the ground state undergoes a series of crossings as $\alpha$ is varied. As $L\to \infty$ we expect these crossings to become dense. In section \ref{sec:alphafamily} we use the Bethe ansatz to give a more precise characterisation of these ground-state level crossings. 

\section{Unified picture through transfer matrices and quantum groups}
\label{sec:transfermatrices}

We have shown that the self-dual $U(1)$ clock models possess a slew of exact degeneracies owing to the presence of the non-abelian Onsager algebra as a symmetry. Moreover, numerical evidence suggests that the structure of the degeneracies is very simple. To make further progress, we exploit the models' integrability. Since a $U(1)$ symmetry is present, the coordinate Bethe ansatz is applicable, and we pursue this approach in section \ref{sec:CBA}. This analysis allows us to understand how the degeneracies are described within the Bethe-ansatz framework, as well as much physical information about the model. The coordinate Bethe ansatz, however, still does not allow us to fully understand the multiplet structure.

Thus before implementing the Bethe ansatz, we describe how to set our Hamiltonians and symmetry algebras in a deeper approach commonly used in integrable models. This approach requires constructing a family of commuting transfer matrices, of which the Hamiltonians are recovered in a particular limit. We show that not only is this possible for the self-dual Hamiltonians $H_n$, but construct transfer matrices for their chiral parts $H_{\rm L}$ and $H_{\rm R}$. Even more strikingly, we can find a transfer matrix that gives a generating function for the elements of the Onsager algebra. We find these transfer matrices by utilising various types of representations of quantum-group algebras \cite{Gomez}. This has a side benefit of giving a nice interpretation of some representations not commonly arising in physics.

\subsection{Correspondence with the higher spin XXZ chains}
\label{sec:XXZ}

A useful starting point is to show how our Hamiltonians can be recast as higher-spin XXZ chains with highly fine-tuned (but still nearest-neighbour) interactions that make them integrable. The connection of the Onsager algebra with such chains has long been known \cite{Roan,Dasmahapatra}, but we provide a direct connection here. Not only does this recasting make finding the corresponding transfer matrix straightforward, but gives insight into how these particular models are special.


In the form \eqref{HnchiralSpm}, the Hamiltonians $H_n$ are not symmetric under spatial parity, whereas the XXZ models are. Parity symmetry (up to boundary conditions) is restored by change of basis  
\be 
H_n =  \left( U_1 U_2 \ldots U_L \right)^{-1}  \widetilde{H}_n  \left( U_1 U_2 \ldots U_L\right) \,,\quad\quad
U_j \equiv e^{ij\pi \left(1+\frac{1}{n}\right) S_j^z}\ .
\label{basischange}
\ee 
The Hamiltonian in this new basis,
\be
\widetilde{H}_n = 
 - \sum_{j=1}^L  \sum_{a=1}^{n-1} 
 \frac{1}{2\sin \frac{\pi a}{n} } 
\left[
 n (-1)^a \left(  
 \left(  S_{j}^- S_{j+1}^+\right)^a 
+
 \left(  S_{j}^+ S_{j+1}^-\right)^a    
    \right)
  +  
    (n-2a)
 \left(   e^{i \frac{ \pi}{n}} \tau_j\right)^a 
     \right]  \,,
 \label{Hntilde}
\ee
is manifestly parity-symmetric in the bulk, although periodic boundary conditions in the original $H_n$ are now {twisted} as \footnote{While the degeneracies and Onsager algebra symmetry described in the previous sections are tied to the choice of boundary conditions \eqref{twist}, we note in passing that for $n$ odd the model \eqref{Hntilde} with periodic boundary conditions commutes with another version of the Onsager algebra, generated by $Q$ and its dual under the modified duality transformation $\tau_j 
\longrightarrow e^{- i \frac{\pi(n+1)}{n}}
\sigma_j^\dagger \sigma_{j+1}
\longrightarrow 
\tau_{j+1} $, yielding
\be 
\widehat{Q}' 
= 
\sum_{j=1}^L \sum_{a=1}^{n-1}  \frac{(-1)^a}{2 i \sin\frac{\pi a}{n}} (  \sigma_j^\dagger \sigma_{j+1})^a  
 \,.
\nonumber
\ee 
} 
\be 
S_{L+1}^\pm = (-1)^L e^{\pm i\pi L/{n}} S_{1}^\pm   \,. 
\label{twist}
\ee 

For example, for $n=2$, 
\be 
\widetilde{H}_2 =  
\sum_{j=1}^{L}  \left(  
 \sigma_j^+ \sigma_{j+1}^-  +   \sigma_j^- \sigma_{j+1}^+    
    \right) 
    =
     \frac{1}{2}\sum_{j=1}^L  
 \left(  
 \sigma_j^x \sigma_{j+1}^x 
+
   \sigma_j^y \sigma_{j+1}^y    
    \right) 
  \,. 
\label{H2new}  
\ee 
The spin-$1/2$ XXZ Hamiltonian is 
\[H_{{\rm XXZ},1/2}=\widetilde{H}_2+\frac{q+q^{-1}}{2} \sum_{j=1}^{L} \sigma^z_j\sigma_{j+1}^z\ \]
and is integrable for any value of the parameter $q$, with gapless behaviour for $|q|=1$ and gapped otherwise \cite{Baxter82}. The Hamiltonian $\widetilde{H}_2$ therefore corresponds to  $q=e^{i\pi/2}$, a special case often called the XX model, well known to be free-fermionic. 

The integrable spin-1 generalisation of the XXZ chain is found by taking the spin-chain limit of the ``19-vertex'' transfer matrix of \cite{FZ}. Again, the integrable line can be parametrised by $q$ with the same gapless/gapped behaviour \cite{XXZSCFT}. The explicit form, however, is much more complicated here, given that for spin $1$, the operators $(S^\pm)^2$ no longer vanish. It is easy to check though that, when $q=e^{i\pi/3}$, the form simplifies and reduces to $\widetilde{H}_3$ from \eqref{Hntilde}. Although we will not exploit it here, it is worth mentioning that the integrable spin-1 chain possesses a very interesting non-local supersymmetry that results in degeneracies between chains of different $L$ \cite{Hagendorf}. 
This supersymmetry commutes with the Onsager symmetries described here.

Higher-spin XXZ Hamiltonians are found by utilising a procedure called ``fusion'' \cite{Kulish81}. As the name indicates, the idea is very much a generalisation of fusing spin-1/2 representations of the $SU(2)$ algebra to get higher-spin representations.  Here however the representations involved are of the quantum-group algebra $U_q(sl_2)$, a one-parameter deformation of $SU(2)$. This algebra has three generators
$\Ss^+, \Ss^-, \Ss^z$ obeying 
\begin{align}
q^{2\Ss^z} \Ss^\pm  q^{-2\Ss^z} = q^{\pm 2}   \Ss^\pm 
 \,, \qquad\quad
[\Ss^+, \Ss^-] =  
\frac{q^{2\Ss^z}-q^{-2\Ss^z}}{q-q^{-1}}\,\,.
\label{qdef}
\end{align}
The relations reduce to $SU(2)$ when $q\to\pm 1$, but in general are not those of a Lie algebra. 

The representation theory of quantum-group algebras depends substantially on whether or not the parameter $q$ is a root of unity.  The reason is apparent in \eqref{qdef}:  in representations where the eigenvalues of ${\bf S}_z$ are  integer or half-integer like in $SU(2)$, the right-hand-side of the latter relation can vanish for $q^n=\pm 1$ for some integer $n$. For any $q$, there occur spin-$S$ representations with $S$ a non-negative integer or half-integer. These act on a chain of $(2S+1)$-state quantum systems, and the action on a single site with basis states  $\{|m\rangle\}_{m=-S, \ldots S}$ is
\bea 
\Ss^z |m \rangle &=& m |m \rangle \,, \qquad m = -S,\ldots,S 
\label{spinSz} 
\\
\Ss^\pm |m \rangle &=&  \sqrt{ [S+1\pm m] [S \mp m]} | m \pm 1 \rangle \,,
\label{spinS} 
\eea
where we have introduced the usual notation
\[[x] \equiv \frac{q^x-q^{-x}}{q-q^{-1}} \ .\] 
For $q^n=\pm 1$, the representation of spin $n/2$ is reducible, not surprising given that (\ref{spinS}) makes it clear that the action of $\Ss^\pm$ can vanish on all states. This reducibility is familiar in physics in the fusion categories arising in anyons or conformal field theory \cite{MooreReshetikhin}. 

Using various properties of the representation theory of quantum-group algebras makes the construction of integrable higher-spin XXZ Hamiltonians straightforward, although technically intricate \cite{XXZSSogo,XXZSBabu,XXZSKiri}.  We find that 
\be
\widetilde{H}_n  \qquad \longleftrightarrow \qquad \mbox{spin-$\frac{n-1}{2}$ XXZ chain at $q=e^{i\pi/n}$.}
\label{identification}
\ee
Although closed-form expressions for the higher-spin Hamiltonians can be found in \cite{XXZSH}, their limit as $q\to e^{i\pi/n}$ is singular, since many terms vanish there: $\widetilde{H}_n$ is much simpler than for generic $q$. This happens because at this value of $q$, we have on each site the highest-spin irreducible representation, so its tensor products used to construct the Hamiltonian are reducible. We thus demonstrate this correspondence for arbitrary $n$ indirectly below, by showing in section \eqref{sec:CBA} that the Bethe equations are the same for the two models.



\subsection{Transfer-matrix construction}

\begin{figure}
\begin{center}
\begin{tikzpicture}[scale=1.2]
\begin{scope}[yshift=0]
\draw[blue,rounded corners=5pt,line width=1pt] (0.25,0.5) --(0,0.5) -- (-0.25,0.25) -- (0,0) -- (10,0) -- (10.25,0.25) -- (10.,0.5) -- (9.75,0.5); 
\foreach \x in {1,2.5,4,8}
{ \draw[line width=1pt] (\x,-.75) -- (\x,.75);
\draw[fill=white,line width=1pt] (\x-0.35,-.25) rectangle (\x+0.35,.25);
}
\node at (6,-0.5) {\Huge $\ldots$};
\node at (1,-1) {$1$};
\node at (2.5,-1) {$2$};
\node at (4,-1) {$3$};
\node at (8,-1) {$L$};
\draw[white, fill=white,line width=1pt] (9,-.25) rectangle (9.6,.25);
\node at (9.3,0) {\small $e^{i \varphi \Ss^z}$};

\node at (1,0) {\small $\mathcal{L}_1(\lambda)$};
\node at (2.5,0) {\small $\mathcal{L}_2(\lambda)$};
\node at (4,0) {\small $\mathcal{L}_3(\lambda)$};

\node at (8,0) {\small $\mathcal{L}_L(\lambda)$};
\end{scope}
\end{tikzpicture}
\end{center}
\caption{
The transfer matrix for the spin-$S$ XXZ chain. The auxilliary space $\mathcal{A}$ is represented in blue, and is  traced over. This construction allows for twisted boundary conditions when an additional $e^{i \varphi \Ss^z}$ acting on $\mathcal{A}$ is inserted.}
\label{fig:TM}
\end{figure}
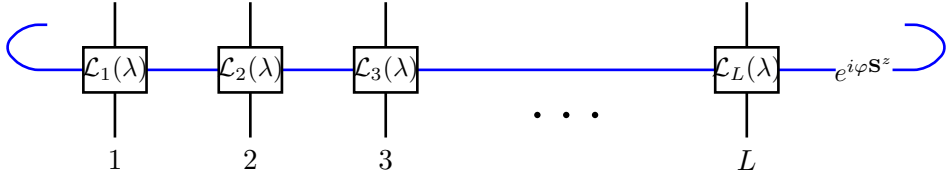

The spin-$S$ XXZ Hamiltonians can be generated from a set of commuting transfer matrices written as \cite{Korepinbook,Gomez}
\be
T(u) = \mathrm{Tr}_\mathcal{A} \left( e^{i \varphi \mathbf{S^z}} \mathcal{L}_L(u) \ldots \mathcal{L}_1(u) \right) \,. 
\label{XXZTu}
\ee
This is pictured in Figure \ref{fig:TM}, with the auxiliary space $\mathcal{A}$ the horitzontal line.
The objects $\mathcal{L}_j(u)$, the so-called Lax operators, are $(2S+1)\times (2S+1)$ matrices acting on the respective sites of the chain, and whose entries are operators  $\Ss^+, \Ss^-, \Ss^z$ acting on $\mathcal{A}$, also $2S+1$-dimensional. 
The trace is over $\mathcal{A}$, and and we have included in \eqref{XXZTu} a factor $e^{i \varphi \Ss^z}$ in order to allow for twisted boundary conditions. To make $\widetilde{H}_n$ periodic we set $\varphi=0$, while to recover the twisted boundary condition in  \eqref{twist} we need to choose 
\be 
\varphi =  L\frac{\pi(n+1)}{n} \,.
\label{twistphi}
\ee 
The simplest case is that of the spin $S=1/2$ chain, where the Lax operators are given by 
\be 
\mathcal{L}(u) =
\left( 
\begin{array}{cc}
[\frac{u}{ \gamma}  +\frac{1}{2}+\Ss^z ]
& 
\Ss^-
\\
\Ss^+
&
[\frac{u}{ \gamma}  +\frac{1}{2}-\Ss^z ]
\end{array}
\right) \,.
\label{lax12}
\ee
The fusion procedure then gives the higher-spin versions \cite{XXZSSogo,XXZSBabu,XXZSKiri, XXZSH}. The Lax operator for $S=1$ is for example derived explicitly in \cite{VernierPiroli}, and is
\bea 
\mathcal{L}(u) =
\left( 
\begin{array}{ccc}
[\frac{u}{ \gamma}  +\Ss^z ]
[\frac{u}{ \gamma}  +1+\Ss^z ]
& 
\Ss^- [\frac{u}{ \gamma}  +\Ss^z ]
&
(\Ss^-)^2 
\\
 \Ss^+ [\frac{u}{ \gamma}  +1+\Ss^z ]
& 
\Ss^+ \Ss^- + [\frac{u}{ \gamma}  +1+\Ss^z ][\frac{u}{ \gamma} -\Ss^z ]
&
\Ss^- [\frac{u}{ \gamma}  -1+\Ss^z ]
\\
(\Ss^+)^2 
&
\Ss^+ [\frac{u}{ \gamma}  -\Ss^z ]
&
[\frac{u}{ \gamma}  +1-\Ss^z ]
[\frac{u}{ \gamma}  -\Ss^z ]
\end{array}
\right)  \,.
\label{lax1}
\eea 

In all cases, the transfer matrices depend on an extra parameter $u$ called the spectral parameter. A fundamental property of the construction is that the transfer matrices associated with different spectral parameters commute with one another:
\be
 [T(u),T(v)]=0  \,.
\ee
One can generate a set of mutually commuting local charges by taking the successive logarithmic derivatives of $T(u)$ about $u=0$, with the Hamiltonian the first one, i.e.
\begin{align}
\widetilde{H}_n = T(0)^{-1} T'(0)\ .
\end{align}

An alternate but equivalent description is to reorganize the matrix elements of the Lax operators into $R$ matrices $R_{\mathcal{A},j}(u)$. These $(2S+1)^2 \times (2S+1)^2$ matrices act on the tensor product of $\mathcal{A}$ with the fundamental representation. The transfer matrix is then
\be
T(u) = \mathrm{Tr}_\mathcal{A} \left( e^{i \varphi \mathbf{S^z}} R_{\mathcal{A}L}(u) \ldots R_{\mathcal{A}1}(u) \right) \,. 
\label{XXZTuR}
\ee
An important property of the $R$ matrices is that at $u=0$, $R_{\mathcal{A},j}(0) \propto \mathcal{P}_{\mathcal{A},j}$, the permutation operator acting as $\mathcal{P}_{\mathcal{A},j} |a\rangle_{\mathcal{A}} \otimes  |b\rangle_j = |b\rangle_{\mathcal{A}} \otimes  |a\rangle_{j}$. It is customary to introduce the matrices  $\check{R}_{\mathcal{A},j}(u) = \mathcal{P}_{\mathcal{A},j} R_{\mathcal{A},j}(u)$, which have the property that $\check{R}(0)$ is proportional to the identity.

\subsection{The chiral Hamiltonians from nilpotent representations}
\label{sec:nilpotent}

The transfer matrices described above are called {\it fundamental}, in the sense that the auxiliary space and the physical sites carry the same spin-$S$ representation. {\it Non-fundamental} transfer matrices are built by using other representations of the quantum-group algebra \cite{Gomez} for the auxiliary space $\mathcal{A}$.  Such transfer matrices have been used extensively in the recent literature on quantum quenches or quantum transport, as generators of quasi-local conserved charges (see e.g.\  \cite{Prosen1,Prosen2,VernierPiroli,DeLuca}). 
The structure of these representations is particularly rich at the points $q^{n}=\pm 1$, where there occur representations with no analog in the $SU(2)$ Lie algebra. At $q^n=-1$, the quantum group $U_q(sl_2)$ has additional $2S+1$-dimensional representations referred to as {\it nilpotent}, {\it semi-cyclic} or {\it cyclic}. Whereas the latter type has arisen previously in studies of the Onsager algebra in the superintegrable chiral Potts model \cite{NishinoDeguchi,Roan}, we describe here how all three arise naturally in our models. The corresponding transfer matrices allow both the chiral Hamiltonians and the Onsager elements to be expressed in an elegant algebraic fashion.

The nilpotent representations are parametrised by a continuous number $\alpha$, as 
\bea 
\Ss^z |m \rangle &=& (m + \alpha ) |m \rangle \,, \qquad m = -S,\ldots,S 
\nonumber
\\
\Ss^+ |m \rangle &=& - [m - S + 2 \alpha] | m + 1 \rangle \,,
\nonumber
 \\
\Ss^- |m \rangle &=&  [m  + S] | m - 1 \rangle \,.
\label{nilpotent}
\eea
Here and from now on we fix $S=(n-1)/2$ and $\gamma = {\pi/n}$, so that $q=e^{i\pi\gamma}$. 
For $\alpha=0$, it is easy to check this reduces the usual spin-$S$ representation \eqref{spinSz}, \eqref{spinS}. Otherwise the representation is non-unitary, and is sometimes referred to as the ``complex-spin representation''.

Using these new generators $\Ss^z, \Ss^+, \Ss^-$ inside the definition of the Lax operator given in the previous section, we now have a two parameter ($u$ and $\alpha$) family of transfer matrices which, crucially, all commute with one another, and so in particular commute with the fundamental transfer matrix \eqref{XXZTu} and the Hamiltonian $\widetilde{H}_n$. 
We label these more general objects as $T(\lambda, \bar{\lambda})$, using the parameters
\be 
\lambda = i u \,, \qquad  \bar{\lambda} = i \gamma \alpha  \,, 
\ee
so that $T(u)=T(-i\lambda,0)$. These transfer matrices obey
\be
[T(\lambda, \bar{\lambda}), T(\lambda', \bar{\lambda}')] = 0 \,.
\ee

Because for the nilpotent representation both the auxiliary space and the physical spins have the same dimension $2S+1$, the logarithmic derivatives with respect to both $\lambda$ and $\bar{\lambda}$ generate independent {\it local} conserved charges. Remarkably, these are the chiral Hamiltonians:
\bea 
  \left. i \frac{\mathrm{d}}{\mathrm{d}\lambda} \log T(\lambda,0) \right|_{\lambda=0}   &=& \frac{2}{n}\widetilde{H}_n =  \frac{2}{n} \left(\widetilde{H}_{\rm R} + \widetilde{H}_{\rm L}\right)   \,,
\label{TH}
 \\
   \left. i \frac{\mathrm{d}}{\mathrm{d}\bar{\lambda}} \log T(0,\bar{\lambda}) \right|_{\bar{\lambda}=0}   &=&  \frac{2}{n} \left(\widetilde{H}_{\rm R} - \widetilde{H}_{\rm L}\right)  
   \,,
\label{THbar}
\eea 
where the tildes arise from the change of basis \eqref{basischange}. Thus the decomposition of $H_n$ into the sum of commuting pieces ${H}_{\rm R}$ and ${H}_{\rm L}$ is expressed very nicely by using an uncommon quantum-group representation. For $n=3$, this fact was known from a classification of three-state quantum chains solvable by coordinate Bethe Ansatz \cite{Ragoucy1,Ragoucy2}, where $\widetilde{H}_{\rm R}$ and $\widetilde{H}_{\rm L}$ are part of a continuous family of Hamiltonians associated with special representations of $U_q(sl_2)$ at roots of unity.

Even more remarkably, the transfer matrices themselves factorise as
\begin{align}
T(\lambda, \bar{\lambda}) =  T(0,0)^{-1} T^{}_{\rm R}(\lambda_{\rm R})  T^{}_{\rm L}(\lambda_{\rm L})  \,,
\label{factorization}
\end{align} 
where $T(0,0)=T(0)$ is the one site translation operator, and where we have introduced 
\begin{align}
 T^{}_{\rm R}(\lambda_{\rm R}) = T\left(\frac{\lambda_{\rm R}}{2},\frac{\lambda_{\rm R}}{2}\right)\ , \qquad\quad
 T^{}_{\rm L}(\lambda_{\rm L}) = T\left(\frac{\lambda_{\rm L}}{ 2}, -\frac{\lambda_{\rm L}}{ 2}\right)  \ .
\end{align}
The transfer matrices generate the chiral Hamiltonians as
\begin{align}
i\left. \frac{\mathrm{d}}{\mathrm{d}\lambda_{\rm R}} \log T_{\rm R}(\lambda_{\rm R}) \right|_{\lambda_{\rm R}=0} = \frac{2}{n}  \widetilde{H}_{\rm R} \ ,\qquad\quad
i\left. \frac{\mathrm{d}}{\mathrm{d}\lambda_{\rm L}} \log T_{\rm L}(\lambda_{\rm L}) \right|_{\lambda_{\rm L}=0}   &= \frac{2}{n}  \widetilde{H}_{\rm L} \ .
\label{HTRL} 
\end{align}
The transfer matrices $T_{\rm R}(\lambda_{\rm R})$ and $T_{\rm L}(\lambda_{\rm L})$ not only form commuting families but commute with one another as well. As our naming indicates, these  are purely chiral, in that their action carries $U(1)$ charge towards the right and the left respectively. The easiest way to prove this is to rewrite the transfer matrix in the $R$-matrix form \eqref{XXZTuR}. In the nilpotent representation, these matrices turn out to be
upper and lower triangular. 

Another important property of nilpotent representations \eqref{nilpotent} is that they are reducible for $\alpha = \pm 1$ (and for any integer value of $\alpha$ not a multiple of $n$). The auxilliary space then is effectively of dimension $n-1$. One can easily check that the action of the generators $\Ss^{z,\pm}$ in the reduced auxilliary spaces for $\alpha=\pm 1$ are equivalent up to a change of basis, so the two corresponding transfer matrices are equal. In terms of $T_{\rm R}$ and $T_{\rm L}$, this translates into the following identity
\be 
T_{\rm L}(\lambda)T_{\rm R}(\lambda+i\pi/n)
=
T_{\rm R}(\lambda)T_{\rm L}(\lambda+i\pi/n) \,, 
\label{TLTRTLTR}
\ee 
as can be checked by direct implementation on the lattice.

\subsection{The Onsager algebra from semi-cyclic representations}
\label{sec:OnsagerTM}

An even more general class of representations of the quantum-group algebra are called (semi-)cyclic.
Transfer matrices built out of these representations do not conserve the $U(1)$ charge but only a ${\mathbb Z}_n$ subgroup.  Moreover, they do not commute with one another, nor in general with the $T(\lambda, \bar{\lambda})$ constructed in the previous subsection. However, under some circumstances \cite{Arnaudon}, one can construct such transfer matrices that do commute with the fundamental one $T(u)=T(\lambda,0)$ with $\lambda=iu$.\footnote{A technical complication is that these circumstances exclude our case $q=e^{i\pi/n}$. A workaround is to use a simple gauge transformation to relate our spin-$S$ XXZ chains to those with $q=-e^{-i\pi/n}$, where the construction works \cite{XXZSKiri}. As a consequence, we can construct transfer matrices for the (semi)cyclic representations for the latter, which after undoing the gauge transformation commute with $\widetilde{H}_{\rm R} + \widetilde{H}_{\rm L}$. } 
Such transfer matrices therefore have precisely the properties of the elements of the Onsager algebra, and we show to find the latter in the former. The connection between the Onsager algebra and cyclic representations of the quantum group has been widely already noted in the past literature, albeit following a different route than the one presented here \cite{Bazhanov}.

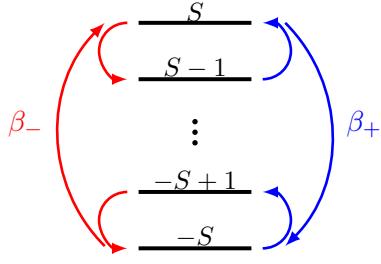
\begin{figure}[ht]
\begin{center}
\begin{tikzpicture}[scale=1.5]
\begin{scope}[yshift=0]
\foreach \x in {0,0.5,1.5,2}
{ \draw[line width=1.5pt] (0,\x) -- (1,\x);
}
\node at (0.5,1.1) {\Huge $\vdots$};
\node at (0.5,0.1) {\small $-S$};
\node at (0.5,0.6) {\small  $-S+1$};
\node at (0.5,1.6) {\small $S-1$};
\node at (0.5,2.1) {\small $S$};

\draw[blue,line width=1pt,-latex] (1.1,0) arc (-90:90:0.25) ;
\draw[blue,line width=1pt,-latex] (1.1,1.5) arc (-90:90:0.25) ;
\draw[blue,line width=1pt,-latex] (1.3,2) arc (45:-45:1.4) ;
\node[blue] at (2,1.1) {$\beta_+$};

\draw[red,line width=1pt,-latex] (-0.1,0.5) arc (90:270:0.25) ;
\draw[red,line width=1pt,-latex] (-0.1,2) arc (90:270:0.25);
\draw[red,line width=1pt,latex-] (-0.3,2) arc (135:225:1.4);
\node[red] at (-1,1.1) {$\beta_-$};
\end{scope}
\end{tikzpicture}
\end{center}
\caption{
Action of the quantum group generators $\Ss^-$ (in red) and $\Ss+$ (in blue) in the (semi)-cyclic representations.
}
\label{fig:cyclic}
\end{figure}

Semi-cyclic representations are characterised by two more parameters, $\beta_{\pm}$. The generators can be written as \cite{Gomez,Prosen2,Arnaudon}
\bea 
\Ss^z |m \rangle &=& (m + \alpha ) |m \rangle \,, \qquad m = -S,\ldots,S 
\nonumber
\\
\Ss^+ |m \rangle &=& \beta_+ \beta_-  + [m-S] [m - S + 2 \alpha] | m + 1 \rangle \,,
\qquad 
\Ss^+ |S \rangle = \beta_+ | -S \rangle \,,
\nonumber
 \\
\Ss^- |m \rangle &=&   | m - 1 \rangle \,,
\qquad 
\Ss^- |-S \rangle = \beta_- | S \rangle \,,
\label{semicyclic}
\eea
in particular for $\beta_+ = \beta_- = 0$ these recover the nilpotent generators \eqref{nilpotent} up to a change of basis.
The action of the generators $\Ss^{\pm}$ in such representations is pictured in Figure \ref{fig:cyclic}. 
The resulting $U(1)$ charge violation and ${\mathbb Z}_n$ preservation is also apparent, in that e.g. ${\bf S}^{\pm}$ in a semi-cyclic representation can change the $U(1)$ charge by $\mp n$.

We define $T_+(\beta_+)$ and $T_-(\beta_-)$ to be the transfer matrices constructed using the semi-cyclic representations with $\beta_-=0$ and $\beta_+=0$ respectively, and $u=\alpha=0$. These commute with $T(\lambda,0)$ and hence $\widetilde{H}_{\rm R} + \widetilde{H}_{\rm L}$, but not with $T(\lambda,\bar{\lambda})$ in general, and so not any $\widetilde{H}(\alpha)$ except at $\alpha=0$ or $\pi$. 
For $n=2$, the R matrix associated with $T_+(\beta_+)$ is
\be 
\check{R}_+(\beta_+) = 
\left(
\begin{array}{cccc}
 1 &   &   &  \\
   & 1  &   &  \\
  &   & 1  &  \\
 \beta_+ &   &   & 1 \\
\end{array}
\right) \,,
\ee 
and similarly that associated with $T_-(\beta_-)$ is obtained by transposing the above expression and replacing $\beta_+$ by $\beta_-$. For $n=3$ we find analogously
\be 
\check{R}_+(\beta_+) = 
\left(
\begin{array}{ccccccccc}
 1 &   &   &  &   &   &   &   &   \\
  & 1 &    &  &   &   &   &   &   \\
  &  & 1  &  &   &   &   &   & \\
  &  & & 1 &  &   &   &      &  \\
  &  &  &  & 1 &   &   &      &  \\
 \beta_+  &  &  &  & & 1 &  &   &   \\
   &  &  &  & &  & 1 &   &   \\
 - \beta_+  &  &  &  &  &  &  & 1 &  \\
  & - \beta_+  &   &  \beta_+  &  &  &  &  & 1 \\
\end{array}
\right) \,, 
\ee 
and similarly for $\check{R}_-(\beta_-)$. 
From there we immediately recognize 
 \begin{align}
\left. \frac{\mathrm{d}}{\mathrm{d}\beta_+} \log T_{+}(\beta_+) \right|_{\beta_+=0} =  2 \sin\frac{\pi}{n} ~ \widetilde{Q}^{+} \ ,\qquad\quad
\left. \frac{\mathrm{d}}{\mathrm{d}\beta_-} \log T_{-}(\beta_-) \right|_{\beta_-=0} =  2 \sin\frac{\pi}{n} ~ \widetilde{Q}^{-} 
\label{QpmT} \,,
\end{align}
which we conjecture to remain true for larger values of $n$. As always, the tilde in \eqref{QpmT} means to take the unitary transform (\ref{basischange}). 

As is typical, higher-order derivatives can be obtained from commutators of the local densities of the first derivatives, namely $\widetilde{Q}^+$ and $\widetilde{Q}^-$ respectively. Since these commutators involve respectively $\Ss^+$ operators only and $\Ss^-$ operators only, the Onsager algebra requires that they commute, and so the higher logarithmic derivatives vanish. Thus our
conjecture for the transfer matrices $T_{\pm}(\beta_\pm) $ can be rewriten as 
\be 
T_{\pm}(\beta_\pm) = T(0) e^{(2 \sin\frac{\pi}{n}) \beta_{\pm} \widetilde{Q}^\pm}  \,.
\ee 

The relations (\ref{THbar},\ref{QpmT}), give the building blocks of the dual $U(1)$ charge, $Q^0$, $Q^+$ and $Q^-$, in terms of the non-fundamental transfer matrices of the higher-spin XXZ quantum chain. All the Onsager elements can be generated by commuting these with each other. It is therefore natural to expect that all the Onsager generators can be expressed in a similar elegant fashion, and we present a conjecture here.

We start with the operators $Q^0_m$, which we refer to as the ``Onsager Hamiltonians'' \footnote{These generators form a subset of the three-parameter abelian subalgebra $I_m = \kappa (A_{m}+A_{-m})+\kappa^* (A_{m+1}+A_{-m+1})+ \mu (G_{m+1}-G_{m-1})$  of the Onsager algebra \cite{Baseilhac18}, corresponding to $\kappa^* = \mu = 0$.}. Since these are mutually commuting, we expect that these are related to the transfer matrices $T_{\rm R}$ and $T_{\rm L}$ constructed from the nilpotent  representation. To this end, we define parameters $\tau_{\rm R}= \tau(\lambda_{\rm R}), \tau_{\rm L}= \tau(\lambda_{\rm L})$ via the function
\be 
\tau(\lambda) = - \tanh \left(\frac{n}{2} \lambda \right) \, .
\label{taudef}
\ee 
We then define a family of commuting local conserved charges as 
\be 
Q_{{\rm R},m} = \frac{1}{ (m-1)!} \left. \frac{\mathrm{d}^m}{\mathrm{d}\tau_{\rm R}^m} \log T_{\rm R} \right|_{\tau_{\rm R}=0}
\,,
\qquad\quad
Q_{{\rm L},m} = \frac{1}{ (m-1)!} \left. \frac{\mathrm{d}^m}{\mathrm{d}\tau_{\rm L}^m} \log T_{\rm L} \right|_{\tau_{\rm L}=0}
\ee
for any positive integer $m$. Generalising the relations for $m=1$ from \eqref{decomposition}, 
we expect that the particular combinations $Q_{{\rm R},m}+Q_{{\rm L},m}$ are in direct correspondence with the higher conserved charges generated by the fundamental transfer matrix \eqref{XXZTu}, while the combinations $Q_{{\rm R},m}-Q_{{\rm L},m}$ are related to the Onsager Hamiltonians. We thus conjecture
\bea 
\widetilde{Q}_{2m+1}^{0} &=&  Q_{{\rm R},2m+1} -  Q_{{\rm L},2m+1}\,, \\
\widetilde{Q}_{2m}^{0} &=&  Q_{{\rm R},2m} -  Q_{{\rm L},2m} - Q  \,.
\eea 
We have checked this conjecture on finite chains for $n=2,3$ and several values of $m$ ranging between $1$ and $10$.

We then consider the formal series expansion
\be
\frac{2\tau(\lambda)}{n} \left. \frac{\mathrm{d}}{\mathrm{d}\bar{\lambda}} \log T(\lambda,\bar{\lambda}) \right|_{\bar{\lambda}=0}
=  \sum_{m=1}^{\infty} {\tau(\lambda)^m}  \widetilde{Q}^{0}_m  - \frac{\tau(\lambda)^2}{1-\tau(\lambda)^2} Q \,, 
\label{gen1}
\ee
 and similarly, using that $\tau(\lambda + i \gamma) = \tau(\lambda)^{-1}$, 
 \be
{2\tau(\lambda)^{-1}\over n} \left. \frac{\mathrm{d}}{\mathrm{d}\bar{\lambda}} \log T(\lambda+i\gamma,\bar{\lambda}) \right|_{\bar{\lambda}=0}
=  \sum_{m=1}^{\infty} {\tau(\lambda)^{-m}}  \widetilde{Q}^{0}_m  - \frac{\tau(\lambda)^{-2}}{1-\tau(\lambda)^{-2}} Q \,. 
\label{gen2}
\ee
 The sum \eqref{gen1}+\eqref{gen2} can be rewritten, after a little rearranging, as the generating function of the Onsager Hamiltonians
 \begin{align}
\mathcal{G}^0(\lambda)  &\equiv
\frac{n}{2i \pi \cosh(n \lambda)} 
  \sum_{p \in \mathbb{Z}} 
 {e^{- |p |\epsilon}}  
  \tau\left(\lambda-i{\gamma/ 2}\right)^p \widetilde{Q}_p^0 \cr 
& =  \frac{1}{2 i \pi} 
 \frac{\mathrm{d}}{\mathrm{d}\lambda}
 \log\left[ 
 {
 T_{\rm R}\left(\lambda-i\frac{\gamma}{2}+i \epsilon\right) 
 T_{\rm L}\left(\lambda+i\frac{\gamma}{2}- i \epsilon\right) 
\over 
T_{\rm L}\left(\lambda-i\frac{\gamma}{2}+i \epsilon\right) 
 T_{\rm R}\left(\lambda+i\frac{\gamma}{2}- i \epsilon\right) 
 }
 \right]   \,.
\label{gen3reg}
 \end{align}
A few comments about this conjecture are in order: first, an operator in the denominator means its inverse has to be taken. Since all of the matrices considered here commute with one another, the notation is unambiguous.
Second, we have introduced in \eqref{gen3reg} a small positive number $\epsilon$, which plays the role of a regulator. In the absence of the latter, the expression \eqref{gen3reg} would vanish as a result of \eqref{TLTRTLTR}.  The interpretation of the regularized generating function \eqref{gen3reg} will become natural in the Bethe-ansatz framework described below.

The remaining Onsager elements can be generated simply commuting with $\widetilde{Q}^{\pm}=\widetilde{Q}^{\pm}_1$ 
as in \eqref{Onsager}. Namely,
 \bea 
\mathcal{G}^+(\lambda) &\equiv&  [\widetilde{Q}^+, \mathcal{G}^0(\lambda)]  =  \frac{n^2}{4\pi \cosh^2(n \lambda)}
  \sum_{p \in \mathbb{Z}} 
 {e^{- |p |\epsilon}}  \tau\left(\lambda+i{\gamma/ 2}\right)^p \widetilde{Q}_p^+ 
\nonumber
 \\
\mathcal{G}^-(\lambda) &\equiv&
 [\widetilde{Q}^-, \mathcal{G}^0(\lambda)]  =
 \frac{n^2}{4\pi \cosh^2(n \lambda)} 
  \sum_{p \in \mathbb{Z}} 
{e^{- |p |\epsilon}}   \tau\left(\lambda+i{\gamma/ 2}\right)^p \widetilde{Q}_p^-  \,,
\label{QplusQminus}
\eea

\section{Bethe-ansatz analysis}
\label{sec:betheansatz}

The existence of a $U(1)$ conserved charge suggests that the energies can be computed using the Coordinate Bethe Ansatz (CBA). This construction will allow us to demonstrate the correspondence with the higher-spin XXZ chains, and provide a means to better understand the structure of the degenerate multiplets.

\subsection{Coordinate Bethe ansatz}
\label{sec:CBA}

The CBA procedure starts with the definition of a reference eigenstate (or pseudovacuum), corresponding to the minimal value of the charge $Q$. Labeling the local basis states for each spin as $n-1,n-2, \ldots 0$, according to the eigenvalue of $(n-1)/2 +Q_j$, the pseudovacuum is defined as 
\be
|\Omega \rangle = |0\ldots 0\rangle
\ee 
From now on we shall shift the Hamiltonians $H_n$ by an appropriate identity term to make $H_n|\Omega\rangle=0$.

\paragraph{One-particle eigenstates}
One-particle eigenstates in the basis \eqref{HnchiralSpm} are plane-wave states 
\be 
| k \rangle =  \sum_{j}  e^{i k x} |1_{j} \rangle
\label{CBA1} \,, 
\ee 
where $|1_{j} \rangle$ stands for the state $|1\rangle$ on site $j$, and $\ket{0}$ on the others. Requiring the periodicity of the wavefunction imposes the quantization $k \in \frac{2\pi}{L} \mathbb{Z}$.
It will be useful to introduce the shifted momenta
\be 
\tilde{k} = k - \frac{\pi(n+1) }{n} \,,
\ee 
which can be understood as the momenta in the basis \eqref{Hntilde}, that is, the momenta for the associated XXZ chains with twisted boundary conditions.
The energy of such states in terms of the latter is easily checked to be given by
\be 
\epsilon(\tilde{k})  =  \frac{n}{ \sin \frac{\pi}{n}} \left(\cos \tilde{k} + \cos \frac{\pi}{n}  \right)
\label{epsilonk}
\,.
\ee 

\paragraph{Two-particle eigenstates}
Two-particle states are given by 
\be 
| k_1,k_2  \rangle =  \sum_{j_1\leq j_2} \left( A_{12} e^{i (k_1 j_1+k_2 j_2)}+ A_{21} e^{i (k_2 j_1+k_1 j_2)} \right) |1_{j_1}1_{j_2} \rangle \,,
\label{psi2}
\ee 
with the convention that $|1_{j}1_{j} \rangle = |2_{j} \rangle$. As follows from examining terms with $j_1$ and $j_2$ far apart, for $|k_1,k_2\rangle$ to be an eigenstate of $H_n$, the energy must be $\epsilon(\tilde{k}_1)+\epsilon(\tilde{k}_2)$. Unwanted terms in $H_n|k_1,k_2\rangle$ with $j_1=j_2\pm 1$ vanish when 
\be 
A_{12} \left(1 +2 \cos \frac{\pi}{n} e^{i \tilde{k}_1} +  e^{i (\tilde{k}_1+\tilde{k}_2)} \right) + 
A_{21} \left(1 +2 \cos \frac{\pi}{n} e^{i \tilde{k}_2} +  e^{i (\tilde{k}_1+\tilde{k}_2)} \right) 
=0 \,.
\label{A12_A21}
\ee 
There are two types of solutions to \eqref{A12_A21}. One is to have both $A_{12}$ and the factor multiplying $A_{21}$ vanish (or the other way around) \cite{Baxtercompleteness}; these so-called $0=0$ solutions will prove pivotal to our discussion. The other way is for them not to vanish, so that
\be
\frac{A_{12}}{A_{21}}= - \frac{1 +2 \cos \frac{\pi}{n} e^{i \tilde{k}_2} +  e^{i (\tilde{k}_1+\tilde{k}_2)}}{1 +2 \cos \frac{\pi}{n} e^{i \tilde{k}_1} +  e^{i (\tilde{k}_1+\tilde{k}_2)} } 
\equiv S(\tilde{k}_1,\tilde{k}_2)
\,.
\label{s_k1_k2}
\ee
Note in particular that $S(\tilde{k},\tilde{k})=-1$, so the wavefunction \eqref{psi2} vanishes when $\tilde{k}_1$ and $\tilde{k}_2$  are equal.
 
Requiring periodicity of the wavefunction then quantizes the momentum via
\begin{align}
e^{iL\pi(n+1)/n} e^{iL\tilde{k}_1}= S(\tilde{k}_1,\tilde{k}_2)\ 
\end{align}
and similarly with $\tilde{k}_1 \leftrightarrow \tilde{k}_2$.

\paragraph{$M$-particle eigenstates}
Nothing in these one- or two-particle eigenstates requires the model to be integrable. However, for the analogous Bethe ansatz for the eigenstates to work for more particles, the model must be integrable.  This fact is clear for $n=2$, where  the $k_j$ are just the free-fermion momenta. Checking that all unwanted terms vanish is straightforward for $n=3$, but making an explicit check gets increasingly difficult for larger values of $n$, as the number of terms in the Hamiltonian increases accordingly. However we verified by explicit implementation for finite chains the validity of the Bethe-ansatz construction up to 3-particle states for $n=4$.
In the following we will therefore take for granted that the Bethe-ansatz construction holds generally, and will describe the general structure of eigenstates. Equivalently, we can just assume that $H_n$ is indeed the appropriate special case of the XXZ chain, and the result follows, since the fusion procedure guarantees the Bethe ansatz will work for all $n$.
 
The $M$-particle eigenstates are parametrized by a set of pseudomomenta $\{k_1, \ldots k_M\}$ as 
\be 
| k_1,\ldots k_M  \rangle =  \sum_{j_1 \leq \ldots \leq j_M} \left( \sum_{{\cal P} \in  \mathfrak{S}_M} A_{\cal P}  e^{i (k_{{\cal P}_1} j_1+ \ldots+ k_{{\cal P}_M} j_M)}
\right)
|1_{j_1} \ldots 1_{j_M} \rangle \,,
\label{psiM}
\ee 
where the second sum if over permutations of the set $\{1, \ldots M \}$ (also labeled as orderings $p_1, \ldots p_M$). In this notation, a state where a spin takes on a value $2,\ldots, n-1$ is included by taking two successive $j_m$ to be equal, e.g.\ $\ket{2_j}=\ket{1_j1_j}$. For the $n$-state model, only $n-1$ consecutive $j_m$ can be equal. The $U(1)$ charge of the state is by construction $-L(n-1)/2+M$. 

These states (\ref{psiM}) are eigenstates of $H_n$ when the coefficients $ A_{\cal P}$ are related by
\be 
\left(1 +2 \cos \frac{\pi}{n} e^{i \tilde{k}_{p_j}} +  e^{i (\tilde{k}_{p_{j}}+\tilde{k}_{p_{j+1}})} \right) A_{\ldots p_j, p_{j+1} \ldots}
+
\left(1 +2 \cos \frac{\pi}{n} e^{i \tilde{k}_{p_{j+1}}} +   e^{i (\tilde{k}_{p_{j}}+\tilde{k}_{p_{j+1}})} \right) 
  A_{\ldots  p_{j+1},p_j \ldots}
=0  
   \,. 
 \label{CBAM}
\ee 
Note in particular from \eqref{CBAM} that having two coinciding pseudomomenta results in a vanishing of all the coefficients $\mathcal{A}_{\cal P}$, as already noticed above for the two-particle states. The pseudomomenta obey an exclusion principle, as typical in Bethe-ansatz eigenstates. Imposing the periodicity of the wavefunction fixes
\be
A_{p_1,p_2 \ldots p_M} = e^{i L k_{p_1}} A_{p_2 \ldots p_M, p_1} \ .
\label{CBAperiodicity}
\ee 
If {both terms in \eqref{CBAM} are non-vanishing}, combining it with the periodicity relation in a quantization of the pseudomomenta $k_j$ through 
\be 
e^{i L \frac{\pi(n +1)}{n}} e^{i L \tilde{k}_j} =  \prod_{m \neq j}^M S(\tilde{k}_j,\tilde{k}_m) = \prod_{m \neq j}^M - \frac{1  + 2 \cos\frac{\pi}{n} e^{i \tilde{k}_m}  + e^{i (\tilde{k}_j+\tilde{k}_m)}}{1 +2 \cos\frac{\pi}{n} e^{i \tilde{k}_j}  + e^{i (\tilde{k}_j+\tilde{k}_m)}}  \ .
\label{BAEk}
\ee
These coupled polynomial equations, one for each $e^{i\tilde{k}_j}$, are known as the Bethe equations. The energy of the corresponding eigenstate is solely given in terms of their solutions as
$E= \sum_{j} \epsilon(\tilde{k}_j)$, with $\epsilon(\tilde{k})$ given by eq. \eqref{epsilonk}.

It is useful to reparametrize the pseudomomenta in terms of the Bethe roots $\{\lambda_j\}$ as  
\be
e^{i \tilde{k}_j} = \frac{\sinh\left( \lambda_j +  i  \frac{\pi}{2n}(n-1)  \right) }{\sinh\left( \lambda_j -  i   \frac{\pi}{2n}(n-1) \right)}   
\,, \qquad 
e^{2 \lambda_j} = \frac{\sin\left( { \tilde{k}_j\over 2} +     \frac{\pi}{2n}(n-1)  \right) }{\sin\left( {\tilde{k}_j\over 2} -    \frac{\pi}{2n}(n-1)  \right) } \,.
\label{klambda}
\ee
The Bethe quantization equations \eqref{BAEk} read in terms of the latter 
\be 
e^{i L \frac{\pi(n +1)}{n}}\left( \frac{\sinh\left( \lambda_j -  i  \frac{\pi}{2n}(n-1) \right) }{\sinh\left( \lambda_j +  i  \frac{\pi}{2n}(n-1) \right)}  \right)^L   = \prod_{l \neq j}^M
\frac{  \sinh\left( \lambda_j - \lambda_l - i \frac{\pi}{n}\right)  }
{  \sinh\left( \lambda_j - \lambda_l + i\frac{\pi}{n}  \right)   }  
\label{BAEl} \,,
\ee 
while the energy becomes
\be 
E =  - \sum_{j=1}^M \frac{n \sin{\pi \over n}}{\cosh(2\lambda_j) + \cos{\pi \over n}} \,.
\label{Energyl}
\ee 
Equations \eqref{BAEl} and \eqref{Energyl} match precisely the Bethe equations and energy of the spin-{$n-1 \over 2$} XXZ chain $q=e^{i\gamma}$ and anisotropy parameter $\gamma = \frac{\pi}{n}$ \cite{XXZSSogo,XXZSBabu,XXZSKiri, XXZSH, XXZSCFT,XXZSCFTFrahm,XXZSCFTDFZ}, up to a rescaling of the energy by a factor $2/n$. This completes the identification of our models with the higher-spin XXZ chains.

\subsection{The degeneracies as exact \texorpdfstring{$n$}{n}-strings}
\label{sec:exactstrings}

As we will now see, the degeneracies described in the beginning of this paper have a very natural interpretation within the Bethe ansatz. 
Degeneracies of this kind have already been studied for the spin-1/2 XXZ chain at $q$ a root of unity, \cite{FabriciusMcCoy,Baxtercompleteness}, and the situation goes much analogously for the case at hand here.  

\begin{figure}
\begin{center}
\begin{tikzpicture}[scale=3.5]
\draw[black] (-1,0) -- (1,0);
\draw[black] (0,-0.6) -- (0,0.8);
\draw[black,dashed] (-1,0.75) node[left] {$\frac{\pi}{2}$} -- (1,0.75);
\draw[black,dotted] (-1,0.25) node[left] {$\frac{\pi}{6}$} -- (1,0.25);
\draw[black,dotted] (-1,-0.25) node[left] {$- \frac{\pi}{6}$} -- (1,-0.25);

\foreach \x in {(-0.679,0.457-0.75),(-0.2946,0.486-0.75),(-0.084,0.4899-0.75),(0.084,0.4899-0.75),(0.2946,0.486-0.75),(0.679,0.457-0.75),
(-0.679,-0.457+0.75),(-0.2946,-0.486+0.75),(-0.084,-0.4899+0.75),(0.084,-0.4899+0.75),(0.2946,-0.486+0.75),(0.679,-0.457+0.75)}
{ \draw[fill=black] \x circle (0.03);
}
\foreach \x in {(-0.8,0),(-0.9,0)}
{ \draw[fill=black] \x circle (0.03);
}
\foreach \x in {(0.2,0.75),(0.5,0.75)}
{ \draw[fill=black] \x circle (0.03);
}
\foreach \x in {(0.42,-0.25),(0.42,0.75),(0.42,0.25)}
{ \draw[red, fill=red] \x circle (0.03);
}

\end{tikzpicture}
\end{center}
\caption{
Example of configuration of Bethe roots for $n=3$. Real roots, ``antistrings'' (of imaginary part $\pi/2$) and 2-strings are shown in black, while an exact 3-string is shown in red.
}
\label{fig:strings}
\end{figure}
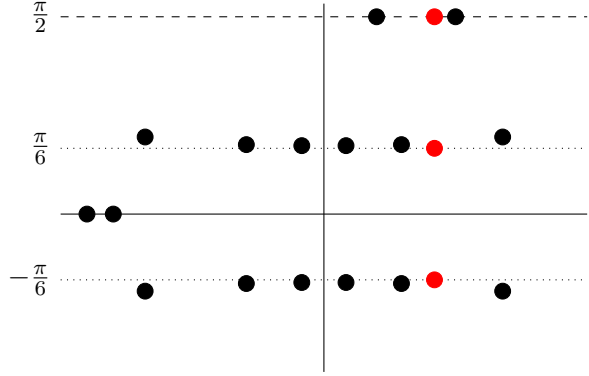

The solutions $\{\lambda_j\}$ of the Bethe equations \eqref{BAEl} typically arrange into sets of real roots, or form patterns in the complex plane. Following a standard argument \cite{takahashi}, the roots assemble into {\it strings}, which are sets of roots distant from one another by approximately $i \gamma$ and centered around the real axis. According to the {\it string hypothesis} \cite{takahashi}, as $L\to\infty$ these values approach 
\be 
\lambda +  \left( j - \frac{p+1}{2} \right) i \gamma \,, \qquad  j=1,\ldots{p}  \,,
\label{pstring}
\ee 
where the real value $\lambda$ is refered to as the {\it string center}. The set above is called a $p$-string. In addition to these, one also encounters the so-called $(1-)$-strings, or antistrings, which are single roots with imaginary part ${\pi \over 2}$ (see Figure \ref{fig:strings} for an illustration).
Strings inherit the exclusion principle verified by Bethe roots, in particular two strings of the same length cannot have the same center. It is a common observation in the study of integrable models that most of the relevant eigenstates of a model, in particular all of its low-energy levels in the large-$L$ limit, are described in terms of the above strings. 
For instance the ground state of the spin-1/2 XXZ chain is described by a set (``sea'') of $L/2$ real Bethe roots (or $1$-strings) on the antiferromagnetic side, and by a set of $L/2$ antistrings (where $\lambda_j+i\pi/2$ is real) on the ferromagnetic side. More generally, the ground state of the spin-$S$ XXZ chain is described by a sea of $2S$-strings on the antiferromagnetic side, and an sea of antistrings on the ferromagnetic side \cite{XXZSSogo,XXZSBabu,XXZSKiri, XXZSCFT,XXZSCFTFrahm}.
The energy associated with a configuration of Bethe roots including strings can be recast in the thermodynamic limit, where the strings become exact, as a sum over the string centers. The contribution to the energy of a $p$-string the form \eqref{pstring} reads for any spin-$S$ XXZ chain
\be 
\lim_{L\to\infty} E_{p \mbox{-}{\rm string}} {=} - \sum_{j=1}^{p} 
\frac{n \sin\gamma}{\cosh(2\lambda + i \gamma(2j-p-1)) + \cos\gamma} \,,
\label{Epstring}
\ee 
and is generically non-zero. 

In these approximate string solutions, the main difference between $q$ a root of unity and $q$ not is that in the former, typically only a finite number of types of string solutions are important as $L\to\infty$. The non-zero energy of (\ref{Epstring}) means that they are not related to the degeneracies in the spectrum. Degeneracies such as we have can arise in the Bethe ansatz from {\it exact $n$-strings}, or {\it exact complete strings}, that occur at root of unity. As the name indicates, the values \eqref{pstring} of the roots are exact even at finite size $L$. Indeed, at the values $\gamma = \frac{\pi}{n}$ the string energy \eqref{Epstring} vanishes for $p=n$. Such string solutions have the form
\be
\{\mu\}_{n} \equiv
\left\{ 
\mu+ i\frac{\pi}{2}  \,,~
\mu + i\frac{\pi}{2} + i \gamma\,,~ 
\ldots 
,
\mu + i\frac{\pi}{2} + i (n-1) \gamma
\right\} \,, 
\label{exactstrings}
\ee 
(see Figure \ref{fig:strings}), where the {\it string center} $\mu$  obeys a slightly different definition than for the ordinary strings above. Since the Bethe roots are defined up to a shift $\lambda_j \to \lambda_j + i \pi$, any cyclic permutation can be performed  within \eqref{exactstrings} so the string center is defined up to shifts by $\pm i \gamma$.
We will also introduce a similar notation, $\{\tilde{k}\}_{n}$ , for the associated (shifted) pseudomomenta, which are related two by two through 
\be 
1 +2 \cos \frac{\pi}{n} e^{i \tilde{k}_{j+1}} +  e^{i (\tilde{k}_j+\tilde{k}_{j+1})}   = 0 \,,\qquad j=1, \ldots, n\,, 
\label{exactstringk}
\ee 
where it is understood that $\tilde{k}_{n+1}=\tilde{k}_{1}$.  In the following, we will also occasionally use the terminology {\it ordinary roots} to denote Bethe roots which are not part of an exact $n$-string. 

In order to understand why exact $n$-strings are special, let us look at their effect on the Bethe equations. Starting from a configuration of roots $\{\lambda_j\}$ solution of \eqref{BAEl}, consider adding to the latter an exact $n$-string of the form \eqref{exactstrings}. The right-hand side of the BAE \eqref{BAEl} for the original roots acquires an additional factor 
\be 
\frac{  \sinh\left( \lambda_j - \mu - i \frac{\pi}{2}- i \frac{\pi}{n}\right)  }
{  \sinh\left( \lambda_j - \mu - i \frac{\pi}{2}+ i\frac{\pi}{n}  \right)   }  
\frac{  \sinh\left( \lambda_j - \mu- i \frac{\pi}{2}  - 2i \frac{\pi}{n}\right)  }
{  \sinh\left( \lambda_j - \mu  - i \frac{\pi}{2}  \right)   }  
\ldots 
\frac{  \sinh\left( \lambda_j - \mu- i \frac{\pi}{2}  - i \pi\right)  }
{  \sinh\left( \lambda_j - \mu  - i \frac{\pi}{2}-(n-2) i\frac{\pi}{n}  \right)   }  
=1  \,, 
\ee 
that is, in terms of the associated pseudomomenta,
\be 
\prod_{\tilde{k}' \in \{\tilde{k}' \}_n} S(\tilde{k}_j,\tilde{k}') = 1   \,.
\label{stringprodS1}
\ee 
so the BAE remain satisfied in the presence of the exact $n$-string. 

Let us turn to the wavefunction, letting $k_1, \ldots k_M$ be the pseudomomenta associated with the original roots $\{\lambda_j\}$, and $\{k_{M+1}\}_n = \{k_{M+1}, \ldots k_{M+n}\}$ those associated with the exact $n$-string. 
The wavefunction associated with $k_1, \ldots k_M$ is given by \eqref{psiM}, where all of the coefficients $A_{\mathcal{P}}$ are fixed by \eqref{CBAM}, up to a global rescaling. 
Upon adding the exact $n$-string, the wavefunction becomes 
\be 
| k_1,\ldots k_{M+n}  \rangle =  \sum_{j_1 \leq \ldots \leq j_{M+n}} \left( \sum_{{\cal P} \in  \mathfrak{S}_{M +n}} A'_{\cal P}  e^{i (k_{{\cal P}_1} j_1+ \ldots+ k_{{\cal P}_{M+n}} j_{M+n})}
\right)
|1_{j_1} \ldots 1_{j_{M+n}} \rangle \,. 
\label{psiMn}
\ee 
 Up to a global rescaling, we can choose 
\be 
A'_{12,\ldots M,M+n, \ldots M+1} = A_{12,\ldots M}
\,,
\ee 
which fixes all of the $(M+n)!$  coefficients $A'_{\mathcal{P}}$ through successive applications of \eqref{CBAM} and of the periodicity condition \eqref{CBAperiodicity}.
Let us look in particular at what happens when two momenta within the exact-string are permuted, say $k_{M+1}$ and $k_{M+2}$. As a consequence of \eqref{exactstringk}, the coefficients must obey
\be 
0 \times  A'_{12,\ldots M,M+n, \ldots M+2,M+1}+ 
  \left(1 +2 \cos \frac{\pi}{n} e^{i \tilde{k}_{M+1}} +   e^{i (\tilde{k}_{M+1}+\tilde{k}_{M+2})} \right) 
  A'_{12,\ldots M,M+n, \ldots M+1,M+2}  = 0\,,
\ee 
which imposes $ A'_{12,\ldots M,M+n, \ldots M+1,M+2}=0$. 
The remaining nonzero coefficients are all obtained from  $A'_{12,\ldots M,M+n, \ldots M+1}$ through permutations made of transpositions within the set of original momenta $k_1, \ldots k_M$ or between the latter and the exact string momenta, but excluding transpositions within the exact string itself. 

The resulting wavefunction, if non-vanishing, is an eigenstate of the Hamiltonian $H_n$ with the same energy as the original $M$-particle state, since the periodicity requirement \eqref{CBAperiodicity} (with $M$ replaced by $M+n$) yields no further constraint than the original Bethe equations obeyed by the momenta $k_1, \ldots k_M$. Moreover, since inserting an exact $N$ string solution includes $n$ more particles, the states with the exact $n$-string has $U(1)$ charge increased by $n$ relative to the corresponding state without it. The exact $n$-strings therefore give degeneracies of exactly the same sort as the Onsager generators do.

While the $M$ original equations are left unaffected by the addition of the exact $n$-string, the additional $M+n$ equations whose left-hand side involves the exact $n$-string itself are ill-defined and do not apply. The center of the exact $n$-string is thus not constrained by the Bethe ansatz. The existence of exact $n$-string solutions can be inferred from the Bethe equations \eqref{BAEl}, as such string solutions make both numerator and denominator vanish. For this reason, exact $n$-strings are sometimes called ``$0/0$'' solutions \cite{FabriciusMcCoy}.  
\footnote{The original motivation of \cite{FabriciusMcCoy} for studying such solutions was to investigate the completeness of the Bethe ansatz, as it was put forward that the Bethe equations fail to uniquely determine states with exact strings. As later argued in \cite{Baxtercompleteness}, this is a mere consequence of the many possible choices of basis vectors within degenerate eigenspaces, and the Bethe ansatz is in fact complete in the sense that it furnishes a complete basis for these degenerate spaces.}

Multiple exact $n$-strings can be added to a given configuration of ordinary roots. The study of the resulting Bethe wavefunction is similar to the preceding. All of the non-zero coefficients are obtained from a reference one through transpositions within the original roots $k_1, \ldots k_M$, between these and momenta of the exact strings, between two momenta in the different exact strings, but not within the exact strings themselves. The sets of equations obtained from the periodicity of the wavefunction once again amount to the original set of equations \eqref{BAEk} for the momenta  $k_1, \ldots k_M$, and no other constraint than the exclusion principle advocated in the previous section is imposed on the location of the exact $n$-strings. 
It follows from this discussion that exact $n$-strings can be arbitrarily added to a Bethe eigenstate to form new eigenstates. Besides the exclusion principle, exact $n$-strings do not influence each other, neither do they affect the quantization equations for the ordinary roots. They can therefore be used to construct degenerate eigenstates in sectors of charges differing by multiples of $n$, which indeed reproduces the structure observed in section \ref{sec:degeneracies}.

\subsection{Quantizing the exact  \texorpdfstring{$n$}{n}-strings: the example of  \texorpdfstring{$n=3$}{n=3} }

The presence of exact $n$-string solutions of the Bethe equations results in degeneracies between states of charge differing by $Q$. However, nothing in the above construction fixes the number of linearly independent choices of exact strings in a given sector, nor the maximal number of strings that can be added on top of a given eigenstate. In other words, the exact $n$-strings are not quantized by the Bethe equations. 

One way of attacking this problem is to utilise
the chiral decomposition \eqref{decomposition}.
Recalling section \ref{sec:chiraldecomposition}, the Hamiltonians $H_{\rm R}$ and $H_{\rm L}$ commute with $H_n$, so they share the same eigenspaces. However, they lift the degeneracies observed in $H_n$. 
We therefore expect that constructing the coordinate Bethe ansatz for $H_{\rm R}$ or $H_{\rm L}$ individually should impose some kind of quantization condition on the exact strings. 
In this section we will sketch this procedure, quickly specializing to $n=3$ for the sake of simplicity. In section \ref{sec:TQ}, we will describe an alternative derivation valid for general $n$.

The CBA construction for the Hamiltonians $H_{\rm R}$ or $H_{\rm L}$ goes very similarly to that for the full Hamiltonian $H_{\rm R}+H_{\rm L}$. \footnote{For $n=3$ the construction has been presented in \cite{Ragoucy1,Ragoucy2} for a general family of Hamiltonians including $H_{\rm R}$ and $H_{\rm L}$, however the exact $n$-strings were not considered there.}    
The one-particle energies are now given by 
\be 
\epsilon_{\rm L}(\tilde{k})  =  \frac{n}{2 \sin \frac{\pi}{n}} \left(e^{i \tilde{k}} + \cos \frac{\pi}{n}  \right) \,,\qquad 
\epsilon_{\rm R}(\tilde{k})  =  \frac{n}{2 \sin \frac{\pi}{n}} \left(e^{-i \tilde{k}} + \cos \frac{\pi}{n}  \right)
\label{epsilonkLR}
\,,
\ee
so that the sum $\epsilon_{\rm R}+\epsilon_{\rm L}$ recovers the energy \eqref{epsilonk} of the full Hamiltonian. 
For generic sets of pseudomomenta, we recover in the same way as in section \ref{sec:CBA} the equations \eqref{BAEk}, \eqref{BAEl}. 
In terms of the parameters $\{\lambda_j\}$ the energies read
\be 
E_{\rm L} = \sum_{j} \frac{n}{2} \tan \left( i \lambda_j - \frac{\pi}{2n}  \right) \,,
\qquad 
E_{\rm R} =  \sum_j \frac{n}{2} \tan \left(- i \lambda_j - \frac{\pi}{2n}  \right) \,.
\label{ELR}
\ee 
From \eqref{ELR}, we verify that exact $n$-strings indeed come with nonzero (but opposite) left and right energies. 
In order to understand how these are quantized by $H_{\rm R, L}$, we will now specify to $n=3$ and consider the first few-particle states.

\paragraph{One exact string on top of the pseudovacuum}

The 3-particle states are written as 
\be 
|k_1, k_2, k_3\rangle = \sum_{j_1\leq j_2 \leq j_3}\sum_{\mathcal{P} \in \mathfrak{S}_3} A_\mathcal{P} e^{i (k_{\mathcal{P}_1} j_1 + k_{\mathcal{P}_2} j_2 + k_{\mathcal{P}_3} j_3)} |1_{j_1} 1_{j_2} 1_{j_3}\rangle \,.
\ee 
We consider the case where the momenta $k_1, k_2, k_3$ form an exact 3-string, namely they are related through equations \eqref{exactstringk} which read in this case
\be 
1+ e^{i \tilde{k}_2} + e^{i (\tilde{k}_1+\tilde{k}_2)} =0  \,,\qquad 
 1+ e^{i \tilde{k}_2} + e^{i (\tilde{k}_2+\tilde{k}_3)} =0 \,,\qquad
  1+ e^{i \tilde{k}_1} + e^{i (\tilde{k}_3+\tilde{k}_1)} =0  \,.
\label{3stringk}
\ee 
As described in section \ref{sec:exactstrings}, the coefficients $A_\mathcal{P}$ are related two by two through the scattering factors between the $k_j$, which are either zero or infinity. Three of them  vanish, $A_{123} = A_{231}= A_{312} =0$, while the others are related by carrying particles around the system, i.e.\ $A_{abc} = A_{cab} e^{i L k_c}$.

Taking the component of the equation $H_{\rm L} |k_1, k_2, k_3\rangle = E_{\rm L} |k_1, k_2, k_3\rangle$ on the state $|j,j+1,j+1 \rangle$, we obtain 
\be 
\sum_{\mathcal{P} \in \mathfrak{S}_3} A_{\mathcal{P}} e^{i \tilde{k}_{\mathcal{P}_2}}e^{i \tilde{k}_{\mathcal{P}_3}} (1 + 
e^{i (\tilde{k}_{\mathcal{P}_2} + \tilde{k}_{\mathcal{P}_3})}
+e^{i \tilde{k}_{\mathcal{P}_1}}+e^{i \tilde{k}_{\mathcal{P}_2}} 
)
= 0 \,,
\label{CBA3}
\ee 
which, restricting to the non-zero terms and using \eqref{3stringk}, becomes 
\be 
\sum_{\mathcal{P} \in \{ 321, 213, 132 \} } A_{\mathcal{P}} e^{i \tilde{k}_{\mathcal{P}_2}}e^{i \tilde{k}_{\mathcal{P}_3}}  e^{i \tilde{k}_{\mathcal{P}_1}} = e^{i (\tilde{k}_1 + \tilde{k}_2 + \tilde{k}_3)} \left(A_{321} + A_{213} + A_{132} \right) =0 \,,
\ee 
which we can rewrite as 
\be 
 e^{i L k_1} + e^{i L (k_1+k_2)} + e^{i L (k_1+k_2+k_3)} = 0 \,.
\label{quantization3}
\ee 
We obtain the analogous equations from cyclic permutations of $k_1, k_2, k_3$, but these are equivalent. Combining them with \eqref{3stringk} yields an 
equation for any one momentum, 
\be 
1 + \left(\frac{e^{i \frac{\pi}{n}} }{1 + e^{i \tilde{k}_i}}  \right)^L 
+
\left(e^{i \frac{2\pi}{n}}  e^{-i \tilde{k}_i}  \right)^L
= 0 \,, 
\ee 
that is 
\be 
2 \cos \left(\frac{L \tilde{k}_i}{2} - \frac{L \pi}{n} \right) 
\left(2 \cos \frac{\tilde{k}_i}{2} \right)^L = -1 
\,.
\ee 
This equation indeed provides a quantization for the centre of the exact $3$-string. 
Working with $H_{\rm R}$ instead of $H_{\rm L}$ one would get the same equation multiplied by an overall minus sign, so effectively the same quantization. In contrast, working with $H_{\rm R}+H_{\rm L}$ would result in an equation of the type $0=0$, and hence no quantization.

\paragraph{One exact string + one particle}

We move on to four-particle states of the form 
\be 
|k', k_1, k_2, k_3\rangle = \sum_{j_0\leq  j_1\leq j_2 \leq j_3}\sum_\mathcal{P} A_\mathcal{P} e^{i (k_{\mathcal{P}_0} j_0 + k_{\mathcal{P}_1} j_1 + k_{\mathcal{P}_2} j_2 + k_{\mathcal{P}_3} j_3)} |1_{j_0} 1_{j_1} 1_{j_2} 1_{j_3}\rangle \,,
\ee 
where $k_1, k_2, k_3$ form an exact string as in the previous paragraph, while $k' \equiv k_0 \in 2\pi \mathbb{Z}/L$ is a solution of the single particle quantization.  
Taking the component of the equation $H_{\rm L} |k', k_1, k_2, k_3\rangle = E_{\rm L} |k', k_1, k_2, k_3\rangle$ on the state $|i,j,j+1,j+1 \rangle$ with $i$ and $j,j+1$ far apart, we get
\be 
\sum_{\mathcal{P} \in \mathfrak{S}_4} A_{\mathcal{P}}e^{i k_{\mathcal{P}_2}}e^{i k_{\mathcal{P}_3}} (
e^{i k_{\mathcal{P}_0}}+e^{i k_{\mathcal{P}_1}}+e^{i k_{\mathcal{P}_2}}+e^{i (k_{\mathcal{P}_2}+k_{\mathcal{P}_3})}+1)
= 0\,.
\ee 
Once again the coefficients $A_{\mathcal{P}}$ vanish for 12 of the 24 permutations, and the remaining 12 are all related to one another through \eqref{CBAM} and \eqref{CBAperiodicity}.
Making similar manipulations as in the above paragraph, we arrive at 
\be 
 e^{i L k_1} S(\tilde{k}',\tilde{k}_1) + e^{i L (k_1+k_2)} S(\tilde{k}',\tilde{k}_1)S(\tilde{k}',\tilde{k}_2)   + e^{i L (k_1+k_2+k_3)} = 0 \,, 
\label{quantization4}
\ee 
which once again yields a quantization of the exact string.

\paragraph{One exact string + $M$ particles}

From \eqref{quantization3}, \eqref{quantization4}, we can conjecture that the quantization equation for an exact 3-string on top of a general background of $M$ other particles $\{k'_j\}$ should read 
\be 
 e^{i L k_1} 
\prod_{j=1}^M S(\tilde{k}'_j,\tilde{k}_1) 
 + e^{i L (k_1+k_2)} \prod_{j=1}^M S(\tilde{k}'_j,\tilde{k}_1) S(\tilde{k}'_j,\tilde{k}_2) + e^{i L (k_1k_2+k_3)} 
=0 \,.
\label{quantization5}
\ee
In section \ref{sec:TQ} we will recover this formula (and generalize it to other values of $n$) through another approach utilising the transfer matrix.  
A particularly remarkable feature is that the quantization of a given exact $n$-string is affected by other particles, but, due to \eqref{stringprodS1}, not by the presence of other exact $n$-strings. In other words, exact $n$-strings do not interact with one another.

We close this discussion by mentioning another proposal for quantizing the exact strings by using a limiting procedure in the anisotropy parameter $\gamma$ \cite{FabriciusMcCoy2}. The two quantizations fundamentally differ in that, while ours should have no relation with the eigenstates at neighbouring values of $\gamma$, that of \cite{FabriciusMcCoy2} should have no relation with the chiral structure $H_{\rm R}, H_{\rm L}$, nor with the underlying Onsager algebra. The two schemes give different results, but we stress that there is no reason why these should coincide: as far as the Hamiltonians $H_n = H_{\rm R} + H_{\rm L}$ (or, equivalently, $\widetilde{H}_n$) are concerned, any quantization of the two strings gives an equally legitimate eigenstate.

\subsection{The quantization equation of exact \texorpdfstring{$n$}{n}-strings, and its solutions}
\label{sec:quantnstrings}

It is quite natural to expect, as will be recovered in section \ref{sec:TQ} through another approach, that equation \eqref{quantization5} should extend to generic values of $n$ as 
\be 
\sum_{m=1}^{n} ~
\prod_{1 \leq j \leq m} 
\left( 
 e^{i L k_j}
 \prod_{p=1}^M S(\tilde{k}'_p, \tilde{k}_j) 
\right) 
=0
 \,,
\label{quantizationTQk}
\ee
where $\{k_1\}_n = {k_1, \ldots k_n}$ denote the pseudomomenta within an exact $n$-string, while $\{k'_p\}_{p=1,\ldots M}$ are the remaining particles on top of which the exact $n$-string is quantized. 
Using the notation \eqref{exactstrings} for the exact $n$-string and denoting by $\{\lambda_k \}_{k=1,\ldots M}$ the Bethe roots associated with the exterior particles, we can rewrite \eqref{quantizationTQk} as 
  \be 
\sum_{m=0}^{n-1} ~
\prod_{1 \leq j \leq m} 
\left[
 \left(e^{i \frac{\pi(n+1)}{n}}
 \frac{\sinh\left(\mu + i\frac{\pi}{2} + i j \gamma + i S \gamma  \right)}{\sinh\left(\mu + i\frac{\pi}{2} + i j \gamma - i S \gamma  \right)} \right)^L
 \prod_{k=1}^M 
 \frac{\sinh\left(\mu + i\frac{\pi}{2} + i j \gamma - \lambda_k-i \gamma  \right)}{\sinh\left(\mu + i\frac{\pi}{2} + i j \gamma - \lambda_k+ i  \gamma  \right)}
 \right]
=0
\label{quantizationTQl}
 \,,
\ee
or alternatively 
\[
\sum_{m=0}^{n-1} ~
 \left(
 \frac{e^{i m \frac{\pi}{n}}\sinh\left(\mu + i\frac{\gamma}{2}  \right)}{\sinh\left(\mu + i\frac{\gamma}{2} + i m \gamma  \right)} \right)^L
 \prod_{k=1}^M 
 \frac{\sinh\left(\mu + i\frac{\pi}{2}  - \lambda_k \right)\sinh\left(\mu + i\frac{\pi}{2}  - \lambda_k +i \gamma  \right)}{\sinh\left(\mu + i\frac{\pi}{2} + i k \gamma - \lambda_k  \right)\sinh\left(\mu + i\frac{\pi}{2} + i (k+1)\gamma - \lambda_k  \right)}
=0
 \,. 
\]
These equations are what we denote as the quantization equations for exact $n$-strings. Since their form is unaffected by the presence of other exact $n$-strings within the set of exterior Bethe roots $\{\lambda_k\}_{k=1, \ldots M}$, we shall assume in the following that $\{\lambda_k\}_{k=1, \ldots M}$ are all ordinary roots, namely contain no exact $n$-string. 

Let us start with the case where there are no background roots, namely $M=0$. The quantization equation \eqref{quantizationTQl} can be rewritten as a polynomial equation of degree $L(n-1)$ in the variable $e^{2 \mu}$, which has therefore $L(n-1)$ zeroes. 
We can check numerically that $e^{2\mu}=0$, that is $\mu = -\infty$, is a zero of \eqref{quantizationTQl} with multiplicity $m_{-\infty}=n-\left( L - n\left\lfloor \frac{L-1}{n} \right\rfloor \right)$.
Such zeros do not correspond to solutions for exact $n$-strings: since these correspond to $\mu \to -\infty$, all $n$-roots of an exact string built out of these would be indistinguishable from one another, and as a result of the exclusion principle between Bethe roots the associated wavefunction would vanish. 

\begin{figure}
\begin{center}
\includegraphics[scale=0.6]{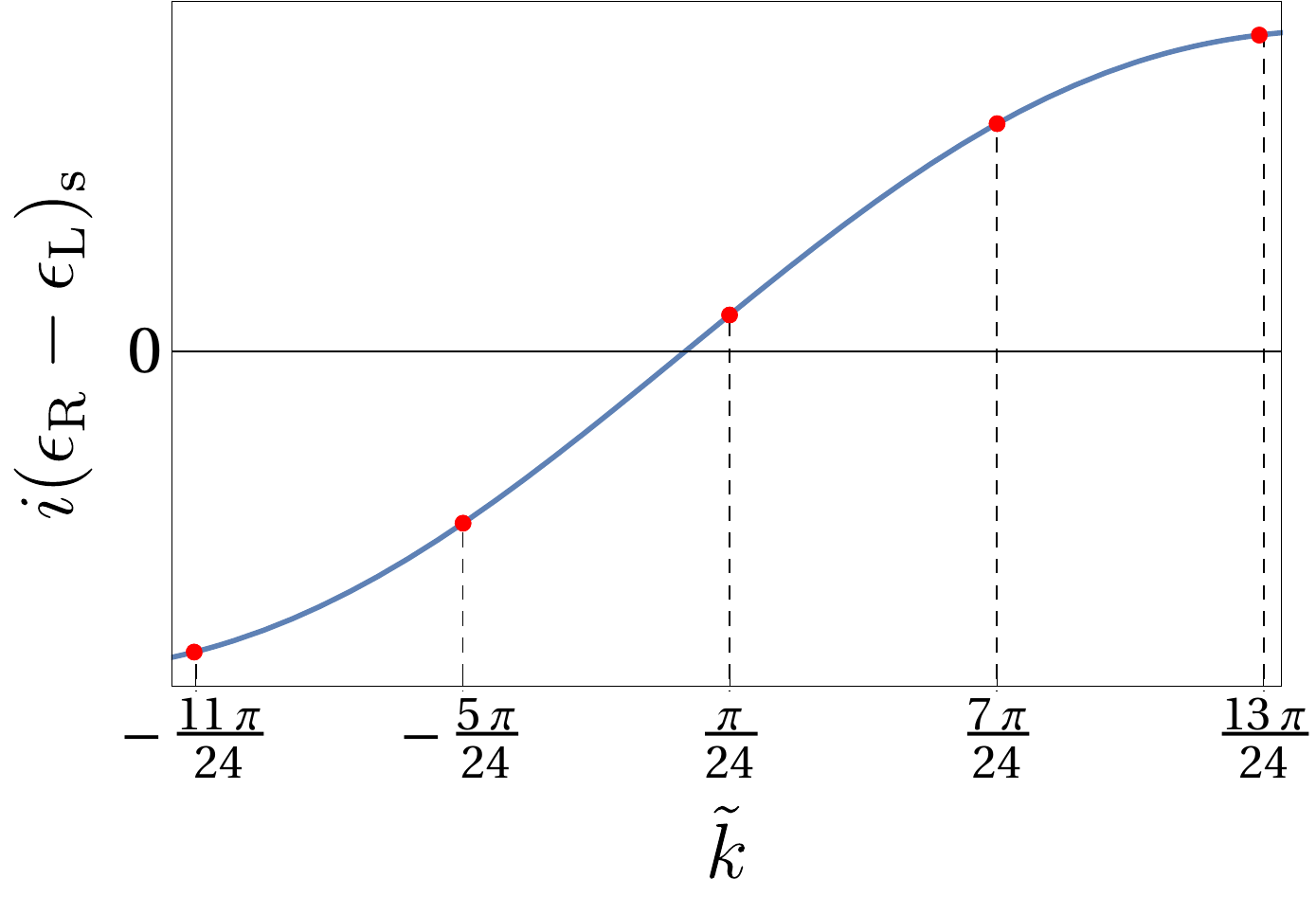}
\end{center}
\caption{
Solutions of the exact $n$-strings quantization equation in absence of other roots for $n=3$, $L=8$. The solutions are represented as red dots, and we have indicated in comparison the values $k \in \pi(2\mathbb{Z}+1)/L$. The blue curve represents the associated energy for the maximally chiral Hamiltonian $i (H_{\rm R}-H_{\rm L})$, whose expression in terms of the associated string centers $\mu$ is given by \eqref{estring}.}
\label{fig:solutions}
\end{figure}

We therefore focus on the remaining finite zeroes. The number of such is a multiple of $n$, namely
\be 
(n-1)L - m_{- \infty} = n\left( L - \left\lfloor(L-1)/ n \right\rfloor -1  \right) \,. 
\ee 
Furthermore, we can check from \eqref{quantizationTQl} that if $\mu$ is a solution, then $\mu+i\gamma$ is a solution. Therefore, the finite zeroes of \eqref{quantizationTQl} form a set of $\left( L - \left\lfloor \frac{L-1}{n} \right\rfloor -1  \right)$ distinct ``exact $n$-strings'', which we parametrize by the set of their centers ${\cal S}$ as 
\be
\{ \mu_k, \mu_k+i\gamma, \ldots \mu_k + i (n-1)\gamma  | \mu_k \in \cal S \} 
\label{eq:solutionS}
\ee 
Note that ${\cal S}$ is defined up to permutations within each of the exact strings. However, we can check numerically that the solutions of \eqref{quantizationTQl} all have $\rm{Im}\, \mu \in \gamma \mathbb{Z}$, so by convention we can define ${\cal S}$ as the set of all real centers. 
For illustration, we represent on figure \ref{fig:solutions} the associated pseudomomenta $\tilde{k}$ for $n=3$ and $L=8$, as well as the associated energy $i (\epsilon_{\rm R} - \epsilon_{\rm L})_{\rm s}$ for the maximally chiral Hamiltonian, \eqref{estring} (see next section).   
A striking feature of these solutions is their proximity to the values corresponding to $k \in \pi(2\mathbb{Z}+1)/L$, reminiscent of a free-fermionic problem such as the one treated in section \ref{sec:n2}. 

With a background of $M$ other particles, the quantization equation can now be rewritten as a polynomial equation of degree $(L+M)(n-1)$ in the variable $e^{2 \mu}$. In fact, the degree is reduced when \eqref{quantizationTQl} has zeroes as $\mu \to \infty$. We checked numerically that the multiplicity of these zeroes is given by $m_{\infty} = n-\left( M - n\left\lfloor \frac{M-1}{n} \right\rfloor \right)$, which results in reducing the degree of the polynomial equation to $(L+M)(n-1)- m_{\infty}$. 
We now turn to the zeroes at $\mu \to -\infty$, that is $e^{2 \mu} = 0$. By numerical inspection, we see that their multiplicity is  $m_{-\infty}=n-\left( L+M - n\left\lfloor \frac{L+M-1}{n} \right\rfloor \right) $. 
Out of the remaining finite zeroes, we checked that for each of the $M$ roots $\lambda_k$ there is a zero at $\lambda_k$, as well as $n$ zeroes corresponding to the exact $n$-string built from $\lambda_k$.  Once again, the exclusion principle implies that neither of these should be considered as solutions for the exact $n$-strings. 

Putting everything together, the number of remaining zeroes is then simply 
\[
(L+M)(n-1) - m_{\infty}-m_{-\infty}-(n+1)M  =
n \left( L - n\left \lfloor  \frac{M-1}{n}  \right\rfloor  
- n\left \lfloor  \frac{L+M-1}{n}  \right\rfloor
-2  \right) \equiv nm_{\cal S}\,.
\]
Once again this is a multiple of $n$, giving  $m_{\cal S}$ distinct exact $n$-strings of the form \eqref{eq:solutionS}. In contrast with the case of no exterior particles, however, we observe that their imaginary parts are not always of the form $\rm{Im} \mu \in \gamma \mathbb{Z}$, so the associated centres cannot always be chosen real. 

Let us now set some notations for the following. For a given eigenstate, we have defined as $\cal{S}$ the set of solutions of the string quantization equation. This equation is unchanged if the considered eigenstate contained exact $n$-strings in the first place, and we therefore define 
\be 
{\cal S} = \rm{s} ~\cup ~ \bar{\rm s} \,,
\label{Ssbar}
\ee 
where $\rm{s}$ and $\bar{\rm s}$  denote respectively the set of  occupied/vacant solutions.

Looking back at the structure of degeneracies detailed in section \ref{sec:degeneracies}, it is now clear where the binomials of equations \eqref{binomials}, \eqref{binomials2} come from. For a given ``highest weight'' state corresponding to a certain configuration of Bethe roots, the exact string quantization equation gives rise to $m_{\cal S}$ solutions. There will be therefore $2^{m_{\cal S}}$ degenerate states, corresponding to the possibilities of having each of these soltions occupied, or empty. Moreover within a sector of fixed charge, that is with a fixed number of exact strings $k$, there will be as many degenerate states as ways to choose $k$ solutions out of $m_{\cal S}$ to be occupied. 
In particular, the ground states of the Hamiltonians $\pm H_{n}$, which are associated with $M=S L$ Bethe roots, have $m_{\cal S}=0$ and are therefore non-degenerate.

\subsection{The ground states of the chiral Hamiltonians \texorpdfstring{$H(\alpha)$}{} }
\label{sec:alphafamily}

Armed with this understanding of the Bethe-ansatz structure, we can return to the family $e^{i \alpha} H_{\rm R} + e^{-i \alpha} H_{\rm L}$, and explore the low-energy physics as $\alpha$ is varied from $0$ to $\pi$.

The physics of the Hamiltonians $\pm H_n$ has previously been explored in their incarnations as higher-spin XXZ chains \cite{XXZSSogo,XXZSBabu,XXZSKiri, XXZSH, XXZSCFT,XXZSCFTFrahm,XXZSCFTDFZ}.
The antiferromagnetic Hamiltonian $H_n$ has a ground state known to be described by a sea of $L/2$ (non exact) $2S=(n-1)$-strings. The low-lying excitations correspond to making holes close to the edges of this sea, or creating a finite number of other types of strings. For real values of the anisotropy $\gamma$ these are found to be gapless, and described by a conformal field theory (CFT) of central charge 
\be 
c= \frac{3S}{S+1}=\frac{3(n-1)}{n+1} \,.
\label{cantiferro}
\ee 
Aspects of this CFT for the $n=3$ model will be studied in much more detail in \cite{Phasediagram}. 

On the ferromagnetic side, corresponding to $H(\pi)=-H_n$, it is known for $S=1/2$ \cite{XXZ} and $S=1$ \cite{Baranowski} that the ground state is described by a sea of $L S$ antistrings. 
It is then natural to expect the same antistring-sea  ground state holds for all $S$, as can be seen by computing the energies associated to the various configurations of strings. Indeed, sending $\lambda\to\lambda+i\pi/2$ changes the sign of \eqref{Epstring}, and it is easy to check for the ferromagnet that the resulting energies are positive for all $p$-strings, and negative for antistrings. The ground state is therefore obtained by filling in the maximal number of the latter, namely $L S$. Studying the low-lying excitations and extracting the scaling of corresponding energies is a standard Bethe-ansatz calculation which we will not pursue here \cite{Korepinbook}, however it quite clear from there that the $c=1$ CFT description should hold for all $S$, in the regime where $\gamma$ is real. This is corroborated by a numerical study of the $n=4$ model, recovering indeed $c=1$.
  \footnote{
  This value of the central charge, as well as \eqref{cantiferro} on the antiferromagnetic side, holds for the periodic XXZ chains. 
The additional boundary twists in $\pm \widetilde{H}_n$ can be interpreted the CFT language as charges at infinity and result in a lowering of the central charge \cite{XXZ,XXZSCFT}. Note however that for $L\in 2n \mathbb{Z}$ the effect of the twist disappears, and the unscreened central charges are recovered.}
 
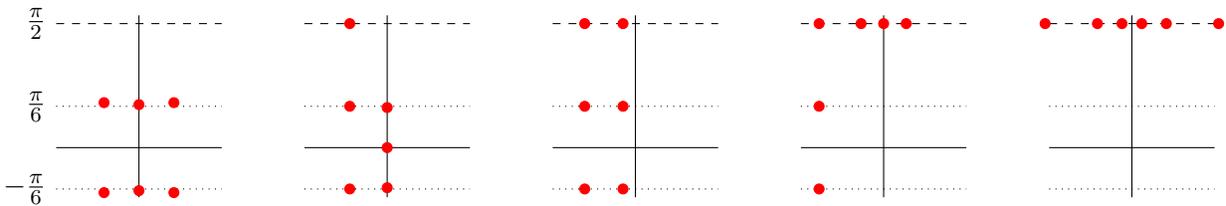
\begin{figure}[ht]
\begin{center}
\begin{tikzpicture}[scale=2.2]

\draw[black] (-0.5,0) -- (0.5,0);
\draw[black] (0,-0.3) -- (0,0.8);
\draw[black,dashed] (-0.5,0.75) node[left] {$\frac{\pi}{2}$} -- (0.5,0.75);
\draw[black,dotted] (-0.5,0.25) node[left] {$\frac{\pi}{6}$} -- (0.5,0.25);
\draw[black,dotted] (-0.5,-0.25) node[left] {$- \frac{\pi}{6}$} -- (0.5,-0.25);
\foreach \x in {(0.211037, 0.27211), (-0.211037, 0.27211), (0., 
  0.260135), (0.211037, -0.27211), (-0.211037, -0.27211), (0., 
-0.260135)}
{ \draw[red,fill=red] \x circle (0.03);
}

\begin{scope}[shift={(1.5,0)}]
\draw[black] (-0.5,0) -- (0.5,0);
\draw[black] (0,-0.3) -- (0,0.8);
\draw[black,dashed] (-0.5,0.75) 
-- (0.5,0.75);
\draw[black,dotted] (-0.5,0.25) 
-- (0.5,0.25);
\draw[black,dotted] (-0.5,-0.25) 
 -- (0.5,-0.25);
\foreach \x in {(0, 0), (0, -0.242476), (0, 0.242476), (-0.226109, 0.75), (-0.226109,0.25), (-0.226109, -0.25)}
{ \draw[red,fill=red] \x circle (0.03);
}
\end{scope}

\begin{scope}[shift={(3,0)}]
\draw[black] (-0.5,0) -- (0.5,0);
\draw[black] (0,-0.3) -- (0,0.8);
\draw[black,dashed] (-0.5,0.75) 
-- (0.5,0.75);
\draw[black,dotted] (-0.5,0.25) 
-- (0.5,0.25);
\draw[black,dotted] (-0.5,-0.25) 
 -- (0.5,-0.25);
\foreach \x in {(-0.0749488, 0.75), (-0.0749488, 0.25), (-0.0749488, -0.25), (-0.306426, 0.75), (-0.306426, 0.25), (-0.306426, -0.25)}
{ \draw[red,fill=red] \x circle (0.03);
}
\end{scope}

\begin{scope}[shift={(4.5,0)}]
\draw[black] (-0.5,0) -- (0.5,0);
\draw[black] (0,-0.3) -- (0,0.8);
\draw[black,dashed] (-0.5,0.75) 
-- (0.5,0.75);
\draw[black,dotted] (-0.5,0.25) 
-- (0.5,0.25);
\draw[black,dotted] (-0.5,-0.25) 
 -- (0.5,-0.25);
\foreach \x in {(0.136342, 0.75), (0, 0.75), (-0.136342, 0.75), (-0.389944, 0.75), (-0.389944, 0.25), (-0.389944, -0.25)}
{ \draw[red,fill=red] \x circle (0.03);
}
\end{scope}

\begin{scope}[shift={(6,0)}]
\draw[black] (-0.5,0) -- (0.5,0);
\draw[black] (0,-0.3) -- (0,0.8);
\draw[black,dashed] (-0.5,0.75) 
-- (0.5,0.75);
\draw[black,dotted] (-0.5,0.25) 
-- (0.5,0.25);
\draw[black,dotted] (-0.5,-0.25) 
 -- (0.5,-0.25);
\foreach \x in {(-0.20822, 0.75), (-0.524302, 0.75), (0.524302, 0.75), (-0.0597185, 0.75), (0.0597185, 0.75), (0.20822, 0.75)}
{ \draw[red,fill=red] \x circle (0.03);
}
\end{scope}

\end{tikzpicture}
\end{center}
\caption{
Configurations of Bethe roots associated with the successive ground states of the Hamiltonian $H(\alpha)$ for $n=3$ on a chain of $L=6$ sites, as $\alpha$ is varied from $0$ to $\pi$. These are the levels highlighted in red in Figure \ref{fig:crossings0}, and correspond to the state of lowest energy in the intervals $\alpha \in [0,\approx 0.484], [\approx 0.484,\approx 0.972], [\approx 0.972,\approx 1.851], [\approx 2.733,\pi]$ respectively. 
}
\label{fig:rootsalpha}
\end{figure}

We now turn to the ``maximally chiral'' Hamiltonian $i (H_{\rm R}-H_{\rm L})$. Recalling \eqref{epsilonkLR}, the single-particle energy is $\tilde{k}$
 \be 
i(\epsilon_{\rm R}-\epsilon_{\rm L})(\tilde{k}) = \frac{n  \sin \tilde{k}}{\sin  \frac{\pi}{n}} = n \tanh \lambda\,.
\ee 
%
The energy of an exact $n$-string, obtained by summing the single-particle energies for all $n$ roots, has a particularly simple form in terms of the string center $\mu$ : 
\be 
i(\epsilon_{\rm R}-\epsilon_{\rm L})_{\rm s}=
n^2 \tanh n\mu \,.
\label{estring}
\ee 
Identifying the Bethe roots associated with the eigenstates of interest by comparing the corresponding energies, we observe that the ground state in that case is comprised solely of exact $n$-strings.
As we have seen in section \ref{sec:quantnstrings}, there are in this case $m_{\cal S} =  L - \left\lfloor \frac{L-1}{n} \right\rfloor -1 $ solutions of the exact $n$-string quantization equations. Those corresponding to $\tilde{k}<0$ (resp. $\tilde{k}>0$)   bring a negative (resp. positive) contribution to the energy, see figure \ref{fig:solutions}. 
As with free fermions, the ground state of $i(H_{\rm R} - H_{\rm L})$ is therefore obtained by filling all the negative energy solutions. The corresponding configuration of exact strings is represented for $n=3$, $L=6$ on the middle diagram of figure \ref{fig:rootsalpha}.  

 As we have described in section \ref{sec:chiralnumerics}, varying $\alpha$ causes the ground state to undergo a series of crossings (see figure \ref{fig:crossings0}). In the language of the Bethe ansatz, increasing $\alpha$ from $0$ to $\pi/2$ corresponds to progressively emptying the sea of (approximate) $n-1$-strings and filling the sea of exact $n$-strings, with some occasional marginal extra roots ensuring that the total number of roots remains the same. 
Similarly, moving from $\alpha = \frac{\pi}{2}$ to $\alpha = \pi$ the crossings result from emptying of the sea of exact $n$-strings and the filling of the sea of antistrings. This is illustrated in figure \ref{fig:rootsalpha}, where we display the Bethe roots associated to the successive ground states for $n=3$, $L=6$. 
It is clear from this mechanism that the number of crossings increases linearly with $L$, and can be expected to become dense throughout the interval $\alpha \in [0,\pi]$ as $L \to \infty$. 

\section{Quantizing the exact  \texorpdfstring{$n$}{n}-strings using the transfer matrix}
\label{sec:TQ}

\subsection{The T-Q relations}

Within the quantum integrability framework, much can be learned from operatorial relations (or functional relations, when viewed at the level of eigenvalues) satisfied by the transfer matrices. A particularly important set are the $T$-$Q$ relations \cite{BaxterTQ} giving the transfer matrices in terms of Baxter's $Q$ operator. The latter has by construction eigenvalues on Bethe states
\begin{align}
 Q(\lambda) =  \prod_j \sinh(\lambda - \lambda_j) \,.
 \label{qfunction}
\end{align}
The $T$-$Q$ relations for the fundamental transfer matrices of the spin-$S$ XXZ chains can be derived from fusion of the spin-1/2 case. 
In the following, we will be interested in $T$-$Q$ equations for the transfer matrices based on nilpotent auxilliary representations at root of unity. Such relations were presented in the case of the spin-1/2 chain in \cite{KorffTQ}, and applications to the study of exact strings can be found in \cite{Korffstrings}. More recent applications of such relations, in particular to the study of quantum quenches and quantum transport, can be found in \cite{DeLuca}.

 It is easy to extend these relations to the  case at hand here, namely spin-$n-1 \over 2$ chains with twisted boundary conditions. We write 
\be 
T(\lambda,\bar{\lambda}) = 
Q\left(\lambda-\bar{\lambda} - i S \gamma \right) 
Q\left(\lambda+\bar{\lambda} + i(S+1) \gamma \right) 
\sum_{m=-S}^{S} e^{i m \varphi}
\frac{ f\left( \lambda+\bar{\lambda} + i\left( m + 1 /2\right)\gamma \right)  }
{
Q\left(\lambda+\bar{\lambda} + i(m+1)\gamma  \right) Q\left(\lambda+\bar{\lambda} + i m \gamma  \right) 
} \,,
\label{TQ}
\ee 
which has exactly the same form as that proposed in \cite{DeLuca}, the only differences residing in the definition of the source function, which is here
\bea 
f(\lambda)  &=& \left(\prod_{k=1}^{n-1} -i  \sinh\left( \lambda +i \left( k-\frac{n}{2}\right)  \gamma  \right)\right)^{L}  
= \left( \frac{1}{2^{n-1}} \frac{\sinh\left(n\left( \lambda - i \frac{\pi}{2} \right)\right)}{\sinh\left( \lambda - i \frac{\pi}{2} \right)}  \right)^L \,,
\eea 
as resulting from the fusion of spin-1/2 chains into a spin-$S$ chain, as well as in the introduction of twist factors in front of each term (we recall that the case of interest for us is obtained to setting the twist parameter as in eq. \eqref{twistphi}).  

 Equation \eqref{TQ}, though not proved, can be checked extensively against exact diagonalization, using the following method adapted from \cite{FabriciusMcCoy2}: since the transfer matrices $T(\lambda, \bar{\lambda})$, whose entries are trigonometric polynomials in $\lambda$ and $\bar{\lambda}$, share the same set of eigenvectors, it is straightforward to show that their eigenvalues are also trigonometric polynomials in $\lambda$ and $\bar{\lambda}$. For a given eigenvector obtained from exact diagonalization of one of these transfer matrices, we can construct this polynomial explicitly by acting on this eigenvector with $T(\lambda, \bar{\lambda})$. Assuming a functional dependence of the form \eqref{TQ}, where the number of Bethe roots is fixed by the $U(1)$ charge but where the Bethe roots themselves are unknowns, we can write from  \eqref{TQ} a trigonometric polynomial equation which should vanish for any $\lambda$, $\bar{\lambda}$. This imposes the cancellation of each coefficient in this equation, which, if a solution exists, fixes the Bethe roots $\{\lambda_j\}$ and confirms the consistency of \eqref{TQ}. 
Using this procedure, we have indeed checked the validity of \eqref{TQ} on finite size chains ($L=4,5,6)$, for several values of $n$, and for various eigenstates in different charge sectors. 

A first observation to make from \eqref{TQ} it can be factorized in the form \eqref{factorization} yielding two individual $T$-$Q$ equations for $T_{\rm L}$ and $T_{\rm R}$:
\bea 
T_{\rm L}(\lambda_{\rm L}) &=& \frac{ Q\left(\lambda_{\rm L} -  i S \gamma \right)  }{ Q\left(i  S \gamma \right) }  
\label{TQL}
\\
T_{\rm R}(\lambda_{\rm R}) &=&   
 Q\left( -i S \gamma \right)
Q\left(\lambda_{\rm R} + i(S+1) \gamma \right)
\sum_{m=-S}^{S} 
\frac{ f\left( \lambda_{\rm R} + i\left( m + 1 /2\right)\gamma \right)  }
{
Q\left(\lambda_{\rm R}+i (m+1)\gamma  \right) Q\left( \lambda_{\rm R} + i m \gamma  \right)
}
\label{TQR} \,,
\eea
which once again can be checked numerically. The manifest asymmetry between \eqref{TQL} and \eqref{TQR} might seem surprising, given that $T_{\rm L}$ and $T_{\rm R}$ are related through a global spin-flip operation. The Bethe ansatz, however, is asymmetric due to the choice of a reference state that breaks the spin-reversal symmetry. The two transfer matrices $T_{\rm L}$ and $T_{\rm R}$ can be understood as the two linearly independent solutions of the $T$-$Q$ equation for the fundamental transfer matrix. These are sometimes referred to as $Q_+$ and $Q_-$, or $Q_{\rm R}$ and $Q_{\rm L}$ in the literature \cite{Bazhanov,KorffTQ,Korffequator}.

It is well-known in the usual case how to recover the Bethe equations from the analyticity properties of the $T$-$Q$ relation \cite{BaxterTQ}.  In the present case, we can go further and derive the quantization equation for exact $n$-strings. 
For this sake, let us introduce a few notations.  
As defined in section \ref{sec:quantnstrings}, for a given eigenstate made of ordinary roots ${\rm r} = \{\lambda_j\}$, the exact $n$-string quantization equations gives rise to a set of solutions $\mathcal{S}={\rm s} \cup \bar{\rm s}$, where ${\rm s}$ and $\bar{\rm s}$ denote respectively the set of occupied and vacant solutions. 
We introduce for each set a different $Q$ function, namely 
\begin{align} 
Q_{\rm r}(\lambda) &=  \prod_{\lambda_j \in {\rm r}} \sinh(\lambda - \lambda_j) \,,
\cr
Q_{\rm s}(\lambda) &= \prod_{\mu_k \in {\rm s}} \sinh\left(\lambda-\mu_k - i \frac{\pi}{2} \right) \sinh\left(\lambda-\mu_k - i \frac{\pi}{2} + i \gamma \right)
\ldots 
 \sinh\left(\lambda-\mu_k - i \frac{\pi}{2} + i(n-1)  \gamma \right)
\,,\cr 
Q_{\bar{\rm s}}(\lambda) &= \prod_{\bar{\mu}_k \in \bar{\rm s}} \sinh\left(\lambda-\bar{\mu}_k - i \frac{\pi}{2} \right) \sinh\left(\lambda-\bar{\mu}_k - i \frac{\pi}{2} + i \gamma \right)
\ldots 
 \sinh\left(\lambda-\bar{\mu}_k - i \frac{\pi}{2} + i(n-1)  \gamma \right)\,, 
 \cr
 Q_{\cal S}(\lambda) &= Q_{\rm s}(\lambda) Q_{\bar{\rm s}}(\lambda)  \,,
\end{align}
so in particular Baxter's original $Q$ function \eqref{qfunction} reads 
\be 
Q(\lambda) = Q_{\rm r}(\lambda) Q_{\rm s}(\lambda) \,.
\label{eq:QsQr}
\ee 

Looking at \eqref{TQR}, it is easy to see that the product $Q_{\rm s}\left(\lambda_{\rm R}+ i(m+1)\gamma  \right) Q_{\rm s}\left( \lambda_{\rm R} + i m \gamma  \right)$ in the denominator does not depend on $m$, and therefore 
 \bea
 T_{\rm R}(\lambda_{\rm R})    
 &=& 
 \frac{
  Q\left( -i S \gamma \right)
Q_{\rm  r}\left(\lambda_{\rm R} +i (S+1) \gamma \right)
}
{
Q_{{\rm s}}\left(\lambda_{\rm R} + i \frac{\gamma}{2}\right)
}
\sum_{m=-S}^{S} 
e^{i m \varphi} \frac{ f\left( \lambda_{\rm R} + i\left( m + 1 /2\right)\gamma \right)  }
{
Q_{\rm r}\left(\lambda_{\rm R}+ i (m+1)\gamma  \right) Q_{\rm r}\left( \lambda_{\rm R} + i m \gamma  \right)
}
\nonumber \\
\label{TQRstrings}
 \eea
Following a standard argument \cite{BaxterTQ}, the functions $T_{\rm R}(\lambda_{\rm R})$ and $T_{\rm L}(\lambda_{\rm L})$ are trigonometric polynomials by construction of the transfer matrix, and should therefore have no poles. 
Taking for instance $\lambda \to \mu_k - i \frac{\gamma}{2}$, this  imposes that the sum in \eqref{TQRstrings} should cancel at this value, namely,
\be
\sum_{m=-S}^{S} 
e^{i m \varphi}  \frac{ f\left(\mu_k + i m \gamma \right)  }
{
Q_{\rm r}\left(\mu_k+ i\left(m- \frac{1}{2}\right)\gamma  \right) Q_{\rm r}\left(\mu_k + i\left(m+ \frac{1}{2}\right) \gamma  \right)
}
=0  \,,
\label{quantizationTQ}
\ee 
which fixes a quantization condition on the exact $n$-string center $\mu_k$. 
Dividing  \eqref{quantizationTQ} by $f(\mu_k + i S \gamma)$ and multiplying by $Q_{\rm r}\left(\mu_k  + i  \frac{\pi}{2}\right)Q_{\rm r}\left(\mu_k  + i  \frac{\pi}{2} - i \gamma\right)$, we indeed recover precisely the quantization equation \eqref{quantizationTQl}. We note that similar results have been obtained for spin-1/2 chains in \cite{Korffstrings}.

We can now factorize \eqref{quantizationTQ} in terms of the solutions of the string quantization equation, which have been described in  section \ref{sec:quantnstrings}. 
 By explicitly implementing all the $Q$ functions above for various eigenstates for $n=3,4,5$ and system sizes $L=3,4,5,6,7$, we check that the following factorization holds 
 \be 
\sum_{m=-S}^{S} 
\frac{ e^{i m \varphi} f\left(\mu + i m \gamma \right)  }
{
Q_{\rm r}\left(\mu+ i\left(m- \frac{1}{2}\right)\gamma  \right) Q_{\rm r}\left(\mu + i\left(m+ \frac{1}{2}\right) \gamma  \right)
}
\propto
e^{(m_{-\infty}-m_{\infty})\mu} 
Q_{\cal S}\left(\mu  \right) \,,
  \label{quantizationTQfactor}
\ee
where the multiplicities $m_{\pm\infty}$ have been defined in section \ref{sec:quantnstrings}, and where the symbol $\propto$ indicates a numerical proportionality constant independent of $\mu$. The latter can be determined for instance by comparing the limits $\mu \to \infty$ of the two sides of \eqref{quantizationTQfactor}, and depends on the state under consideration. From there, we can rewrite $T_{\rm R}(\lambda_{\rm R})$ as
\bea 
T_{\rm R}(\lambda_{\rm R}) &\propto & 
\frac{
Q(-i S \gamma) Q_{\rm  r}\left(\lambda_{\rm R} +i (S+1) \gamma \right) }
{
Q_{\rm s}\left(\lambda_{\rm R} + i \frac{\gamma}{2}\right)
}
Q_{\rm s}\left(\lambda_{\rm R} + i \frac{\gamma}{2}\right)
Q_{\bar{\rm s}}\left(\lambda_{\rm R} + i \frac{\gamma}{2}\right)
\nonumber
\\
&\propto & 
Q(-i S \gamma) Q_{\rm  r}\left(\lambda_{\rm R} +i (S+1) \gamma \right) 
Q_{\bar{\rm s}}\left(\lambda_{\rm R} + i \frac{\gamma}{2}\right) \,,
\label{TRconj}
\eea 
which will turn out useful in the following.

\subsection{The exact string creation/annihilation operators}
\label{sec:stringcreation}

We finally are able to give a precise link between the exact strings and the elements of the Onsager algebra. 
To do so, we utilise the generating functions $\mathcal{G}^{0}(\lambda), \mathcal{G}^{\pm}(\lambda)$ introduced in section \ref{sec:OnsagerTM} from the transfer matrix construction.
Consider the commutators of the generating functions $\mathcal{G}^{\pm}(\lambda)$ with the Onsager Hamiltonian $\hat{Q}^0\propto  H_{\rm R}-H_{\rm L}$. Using the commutation relations \eqref{Onsager}, we obtain
\be 
\left[ \hat{Q}^0 , \mathcal{G}^\pm(\lambda)  \right] 
= \pm \frac{n}{2}(\tau\left(\lambda-i{\gamma/ 2}\right) + \tau\left(\lambda+i{\gamma/ 2}\right)) \mathcal{G}^\pm(\lambda) \,, 
\ee  
Using the definition \eqref{taudef} of $\tau$ and the expression \eqref{estring} for the energy of an exact $n$-string then gives,, in the limit $\epsilon \to 0$,
\be 
\left[ i(H_{\rm R}-H_{\rm L}) , \mathcal{G}^\pm(\lambda)  \right] 
\ =\ \pm 
n^2 \tanh(n \lambda)\mathcal{G}^\pm(\lambda)  
\ =\  \pm 
i(\epsilon_{\rm R} - \epsilon_{\rm L})_{\rm s}(\lambda)
\mathcal{G}^\pm(\lambda)  
  \,.
  \label{comGpm}
\ee 
The generating functions $\mathcal{G}^\pm(\lambda)$ thus have the same  commutation relations with $i(H_{\rm R}-H_{\rm L})$ as would operators creating or annihilating an exact $n$-string with center $\lambda$.

In order to make this correspondence more precise, we need to take proper care of the regulator in \eqref{gen3reg}. 
Let us first consider the action of $\mathcal{G}^0(\lambda)$ on a given eigenstate specified by a set of ordinary roots $\{ \lambda_j\}$ and of exact $n-$strings of centers $\{ \mu_k\}$. 
The eigenvalues of $T_{\rm R}$ and $T_{\rm L}$ were expressed in terms of the various $Q$ functions in section \ref{sec:TQ}, and we further recall the alternative conjectured expression \eqref{TRconj} for $T_{\rm R}$, which has been verified through extensive numerical checks. 
From there we obtain 
\be 
{
 T_{\rm R}\left(\lambda-i\frac{\gamma}{2}+i \epsilon\right) 
 T_{\rm L}\left(\lambda+i\frac{\gamma}{2}- i \epsilon\right) 
\over 
T_{\rm L}\left(\lambda-i\frac{\gamma}{2}+i \epsilon\right) 
 T_{\rm R}\left(\lambda+i\frac{\gamma}{2}- i \epsilon\right) 
 }
 =
 \frac{Q_{\bar{\rm s}}(\lambda + i \epsilon)Q_{\rm s}(\lambda - i \epsilon)}
 {Q_{\rm s}(\lambda + i \epsilon)Q_{\bar{\rm s}}(\lambda - i \epsilon)} \,,
\ee 
where we recall that $Q_{s}$ and $Q_{\bar{\rm s}}$ indicate products over the occupied exact $n$-strings, and the vacant exact $n$-strings solutions respectively. We then have 
\be 
{T_{\rm R}\left(\lambda-i\frac{\gamma}{2}+i \epsilon\right) 
 T_{\rm L}\left(\lambda+i\frac{\gamma}{2}- i \epsilon\right) 
\over 
T_{\rm L}\left(\lambda-i\frac{\gamma}{2}+i \epsilon\right) 
 T_{\rm R}\left(\lambda+i\frac{\gamma}{2}- i \epsilon\right) 
 }
 =
 \prod_{\mu \in {\rm  s}}
 \frac{\lambda - \mu - i \epsilon}
 {\lambda - \mu + i \epsilon} 
 \prod_{\bar{\mu} \in \bar{\rm s}}
\frac{\lambda - \bar{\mu} + i \epsilon}
 {\lambda - \bar{\mu} - i \epsilon}  \,,
\ee 
and so for $\epsilon \to 0^+$,
\begin{align}
\mathcal{G}^0(\lambda)  &=
\frac{1}{\pi}
\left(
\sum_{\mu \in {\rm  s}} \frac{ \epsilon}{(\lambda-\mu)^2 + \epsilon^2}
- 
\sum_{\bar{\mu} \in \bar{\rm s}} \frac{ \epsilon}{(\lambda-\bar{\mu})^2 + \epsilon^2}  
 \right) 
 \cr
 &   \stackrel{\epsilon\to 0^+}{\longrightarrow}   
\sum_{\mu \in {\rm  s}} \delta(\lambda-\mu)
-
\sum_{\bar{\mu} \in \bar{\rm s}} \delta(\lambda-\bar{\mu})   \,.
\label{lorentzians}
\end{align}

Consider now $\mathcal{G}^+(\lambda)$ acting on a given eigenstate $|\Psi\rangle$. $\hat{Q}^+ |\Psi\rangle$, if non-vanishing, is degenerate with $| \Psi \rangle$ (with respect to the original Hamiltonian $H_{\rm R}+H_{\rm L}$) and has $n$ more particles, so can be expanded as a combination of all possible exact $n$-strings that can be built on top of $|\Psi\rangle$. In transparent notation, 
\be 
\hat{Q}^+  |\Psi\rangle = \sum_{ \bar{\mu} \in \bar{\rm  s}} \alpha_   {\bar{\mu}} | \Psi \cup \{\bar{\mu}\}_n  \rangle  \,,
\ee 
where $\alpha_{\bar{\mu}}$ are some coefficients. 
From there, 
\begin{align}
\mathcal{G}^+(\lambda) |\Psi\rangle
&=  [\hat{Q}^+, \mathcal{G}^0(\lambda)] |\Psi\rangle 
\cr
&= 
  \sum_{\bar{\mu} \in  \bar{\rm  s}} \alpha_{\bar{\mu}} \delta(\lambda - \bar{\mu})   | \Psi \cup \{\bar{\mu}\}_n  \rangle \,,
\end{align}
Thus $\mathcal{G}^+(\lambda)$ creates an exact $n$-string at center $\lambda$ whenever this is allowed. 
Similarly, for $\mathcal{G}^{-}(\lambda)$ we find
\bea 
\mathcal{G}^-(\lambda) |\Psi\rangle
=
  \sum_{{\mu} \in {\rm s}} \alpha_{{\mu}}  \delta(\lambda -  {\mu} )   | \Psi \setminus \{ {\mu}  \}_n  \rangle \,,
\eea 
so $\mathcal{G}^-(\lambda)$ annihilates an exact $n$-string at center $\lambda$, whenever this is allowed. 
In conclusion, the operators $\mathcal{G}^{\pm}(\lambda)$ are precisely the string creation/annihilation operators!

In order to check the validity of our construction, it is worth having a look at slightly different objects, namely $\epsilon ~\mathcal{G}^{0}(\lambda)$, $\epsilon ~\mathcal{G}^{+}(\lambda)$ and $\epsilon ~\mathcal{G}^{-}(\lambda)$ in  the $\epsilon \to 0$ limit.
As for the previously considered $\mathcal{G}^{0}(\lambda)$, $\mathcal{G}^{+}(\lambda)$ and $\mathcal{G}^{-}(\lambda)$, these are zero for most values of $\lambda$, except when $\lambda$ is a solution of the exact $n$-string quantization equation on top of some state. In that latter case, the Lorentzians in \eqref{lorentzians} have a finite $\epsilon \to 0$ limit when multiplied by $\epsilon$, so  $\epsilon ~\mathcal{G}^{0}(\lambda)$, $\epsilon ~\mathcal{G}^{+}(\lambda)$ and $\epsilon ~\mathcal{G}^{-}(\lambda)$ become well-defined, finite operators. 
As a check, we have constructed the operators $\epsilon ~\mathcal{G}^{\pm}(\lambda)$ explicitly on the lattice, and verified for a few examples that these indeed act as  exact $n$-string creation/annihilation operators. 
In the $n=2$ case in particular, we can verify that these recover the creation and annihilation operators introduced in section \ref{sec:n2}. Consider indeed the formal expansions \eqref{QplusQminus} for $\mathcal{G}^\pm(\lambda)$. In terms of the pseudomomentum $k$ related to $\lambda$, we have for $n=2$ the simple relation 
\be 
\tau\left(\lambda+i{\gamma/ 2}\right) = e^{i \left(k + \frac{\pi}{2} \right)} \,. 
\ee  
From there, the generating functions (which we denote by $\mathcal{G}^{\pm}(k)$ by abuse of notation) read   
\bea 
\mathcal{G}^{\pm}(k) &=& - \frac{2  \cos k}{\pi} 
  \sum_{m=1}^{\infty} 
{e^{- m \epsilon}}   \sin\left(m \left( k + \frac{\pi}{2}  \right) \right) \widetilde{Q}_m^{\pm}   \,.
\eea 
Using the periodicity of the Onsager algebra for $n=2$, $Q_{m+L}^\pm = (-1)^{Q+1} Q_{m}^{\pm}$, the infinite sum is non-vanishing in the $\epsilon\to 0$ limit only when $e^{i k L} = (-1)^{Q+1}$, which is  precisely the quantization equation discussed in section \ref{sec:n2}.
We get in that case 
\bea 
\lim_{\epsilon \to 0} \epsilon ~ \mathcal{G}^{\pm}(k) &=& - \frac{2  \cos k}{\pi} {\lim_{\epsilon\to 0}  \frac{\epsilon}{1-e^{- L \epsilon} }}
  \sum_{m=1}^{L-1} 
  \sin\left(m \left( k + \frac{\pi}{2}  \right) \right) \widetilde{Q}_m^{\pm} \nonumber \\
  &=& - \frac{2  \cos k}{L \pi}  
  \sum_{m=1}^{L-1} 
  \sin\left(m \left( k + \frac{\pi}{2}  \right) \right) \widetilde{Q}_m^{\pm} \,,
\eea
which, up to a proportionality factor and the change of basis \eqref{basischange}, precisely recovers the operators ${\cal Q}(k)$ and ${\cal Q}^\dagger(k)$ of section \ref{sec:n2} (see equation \eqref{SQ}).  
 
The operators $\mathcal{G}^{r}(\lambda)$, in the $\epsilon \to 0$ limit, make sense as distributions. Another way to build finite norm lattice operators out of these is to consider integrals over $\lambda$. 
Looking back at the definition \eqref{gen3reg}, these can be recast as contour integrals of logarithmic derivatives of the transfer matrices $T_{\rm R}$, $T_{\rm L}$. As an example, the ground state of the maximally chiral Hamiltonian $i(H_{\rm R}- H_{\rm L})$, which as we have seen in section \ref{sec:alphafamily} is made purely of exact $n$-strings, could be formally constructed from a product of such operators. It remains unclear, however, how much of a practical use such a construction might be.

\section{Conclusion}

We started the paper by explaining how the Onsager algebra is a symmetry algebra of lattice models with both a $U(1)$ symmetry and self-duality. Such a non-abelian symmetry will result in exact degeneracies in the spectrum, and we described how they appear in an $n$-state clock model. Moreover, in these models the Onsager algebra is intimately related to a quantum-group algebra, providing a nice physical realisation of non-fundamental representations of the latter. Because this model is not free-fermionic, the Onsager algebra cannot be used to compute exactly the symmetry multiplets, so we resorted to a more detailed analysis coordinate Bethe-ansatz analysis and a set of deep functional relations. The symmetry structure admits an elegant description in term of exact $n$-string solutions of the Bethe equations, and we showed how to construct operators annihilating and creating them.

Our work suggests a number of future directions to pursue. The superintegrable chiral Potts Hamiltonian \eqref{HSI} is built from the symmetry generators $Q$ and $\Qh$, and so commutes with our Hamiltonian $H_n$. Since the latter has a $U(1)$ symmetry and so is easily treated using the coordinate Bethe ansatz, it gives a simple and direct way of understanding why the corresponding Bethe equations also arise in the integrable chiral Potts models; previous analyses were somewhat indirect.  Our results therefore may provide some new insight into these models and their integrable structure.

The continuum limits of both $H_n$ and $-H_n$ are described by conformal field theories. This implies that the Onsager algebra survive sin the continuum limit, as the degeneracies not only survive but are further enhanced. Indeed, an infinite-dimensional symmetry algebra, the Virasoro algebra, is a symmetry of all conformal field theories, and those with a $U(1)$ symmetry like ours have an even larger symmetry generated by a Kac-Moody algebra \cite{Goddard86}. Since some of the conformal structure already is apparent on the lattice \cite{Koo93,Zou17}, it likely would be fruitful to examine the connection of the Onsager algebra with these conformal symmetry algebras. Indeed, it is not difficult to see the connection in the $n=2$ free-fermion case. However, the lattice chiral decomposition of the Hamiltonian into commuting left and right-moving parts is not the same as the analogous decomposition in the conformal field theory, since the empty state is not the ground state.  It thus would be quite interesting to understand what happens at higher $n$. 

Even more tantalisingly, our result that the Onsager-algebra symmetry arises from $U(1)$ symmetry and self-duality does not seem to have anything inherently to do with our models being $1+1$ or two-dimensional.  Is it possible for Onsager symmetries to arise in higher-dimensional self-dual models?

\subsection*{Acknowledgments}

E.V thanks Eric Ragoucy and Luc Frappat for clarifications about their work \cite{Ragoucy1,Ragoucy2}, as well as Pascal Baseilhac, Samuel Belliard, Bruno Bertini, Azat Gainutdinov, Barry McCoy, Christian Korff, Giuliano Niccoli and Hubert Saleur for discussions. This work was supported by EPSRC through grant EP/N01930X.

\bigskip

\section*{Appendix:  The manifestly $U(1)$-invariant form of $H_n$}
\setcounter{equation}{0}
\renewcommand{\theequation}{A\arabic{equation}}

The $U(1)$ charge $Q$ from \eqref{Qdef} is a sum over the $\tau$ operators, and so the only non-commuting terms in the Hamiltonian  \eqref{Hnchiral} involve some $(\sigma^\dagger_j\sigma_{j+1})^a$. It is thus useful to rewrite  \eqref{Hnchiral} in the form 
\begin{align}
 H_n = i\sum\limits_{j=1}^L \sum\limits_{a=1}^{n-1} \frac{1}{1-\omega^{-a}} 
 \biggl[ (2a-n) \tau_j^a + \frac{1}{2} \big(\sigma^\dagger_j\sigma_{j+1}\big)^a\,
 \Big({R}^{(n-a)}_{j}-R^{(a)}_{j+1}\Big)\biggr]\ , 
     \label{HR}
\end{align}
where
\[
R^{(a)}_j  =  n-2a - 2\sum_{b=1}^{n-1}\frac{1-\omega^{-ab}}{1-\omega^{-b}}\tau_j^b\ .
\]
The key property of the latter operators is that for $a=0\dots n-1$ their square is proportional to the identity:
$\big(R^{(a)}_j\big)^2 = n^2. $
Their eigenvalues are therefore all $\pm n$. Indeed, in the basis \eqref{tausigma} all the $R^{(a)}_j$ are diagonal, and letting the eigenvalues of the $\tau_j$ be $\omega^{t_j}$ gives 
\begin{align}
R^{(a)}_j = 
\begin{cases}
-n\qquad & t_j=0\dots a-1\ ,\cr
\ \, n & t_j=a\dots n-1\ .
\end{cases}
\label{Rdiag}
\end{align}

To show how the $S^\pm$ appear, note that the operator $\sigma^\dagger_{j}$ shifts $t_j$ by $+1\,$mod$\,n$, while $\sigma_{j+1}$ shifts $t_{j+1}$ by $-1\,$mod$\,n$. Thus $\sigma^\dagger_{j}\sigma_{j+1}$ violates $U(1)$ conservation when acting on the states with either $p_j =n-1$ or $p_{j+1}=0$, but not both. Conveniently, from \eqref{Rdiag} it follows that the linear combination ${R}^{(n-a)}_{j}-R^{(a)}_{j+1}$ annihilates these states, preserving the $U(1)$.
Proceeding in this fashion gives
\[ \big(\sigma^\dagger_j\sigma_{j+1}\big)^a\Big({R}^{(n-a)}_{j}-R^{(a)}_{j+1}\Big)= 2n \big(S^+_jS^-_{j+1}\big)^{n-a}- 2n \big(S^-_j S^+_{j+1}\big)^a\ ,
\]
which when used in \eqref{HR} yield the manifestly $U(1)$-invariant Hamiltonian \eqref{HnchiralSpm}.

\end{document}